# MARS IN THE AUSTRALIAN PRESS, 1875–1899.
# 1. INTERPRETATION, AUTHORITY AND PLANETARY SCIENCE

**Richard de Grijs**
School of Mathematical and Physical Sciences, Macquarie University,
Balaclava Road, Sydney, NSW 2109, Australia
Email: richard.de-grijs@mq.edu.au

**Abstract:** In the late nineteenth century, Mars emerged as one of the most intensively reported astronomical objects in the popular press, driven by favourable oppositions, improved telescopic capabilities and growing speculation regarding planetary habitability. This paper examines how Mars was interpreted in Australian newspapers between the 1870s and 1899, focusing on the ways in which astronomical knowledge was framed, contextualised and debated within a colonial media environment. Drawing on a large collection of digitised newspaper articles from the National Library of Australia, this study analyses how observational authority, instrumental credibility and individual expertise were harnessed in press reporting. The paper situates Australian Mars coverage within a global network of scientific communication dominated by metropolitan centres in Europe and North America, while highlighting the distinctive role played by southern-hemisphere visibility. Australian observatories and observers—most notably Robert J. L. Ellery, Henry Chamberlain Russell and Walter Frederick Gale—were frequently positioned as contributors of confirmatory observation rather than interpretive leadership, reinforcing a pattern of locally grounded but internationally oriented scientific engagement. The analysis traces a shift from early emphasis on disciplined observation and measurement to later periods characterised by contested interpretations, particularly surrounding the so-called Martian "canals" and the speculative claims advanced by personalities such as Percival Lowell in the USA. By examining how newspapers mediated between observational astronomy, engineering analogies and popular imagination, this study contributes to a broader understanding of how planetary science entered public discourse beyond metropolitan centres. In doing so, it underscores the active role of colonial newspapers in shaping scientific meaning and situates Australian Mars reporting within the wider history of nineteenth-century astronomical culture.

## 1. INTRODUCTION

In the final quarter of the nineteenth century, Mars emerged as one of the most easily visible objects of astronomical inquiry. This visibility was closely tied to the planet's periodic *oppositions*—configurations in which Mars and the Sun are located on opposite sides of the Earth, bringing Mars to its closest approach and making it appear brighter and larger than usual in the night sky.

At a favourable opposition, Mars can approach the Earth to roughly 55–60 million km, compared with an average Earth–Mars separation of about 225 million km and a maximum distance exceeding 400 million km (when the two planets are in conjunction, i.e. found on opposite sides of the Sun). Correspondingly, the apparent angular diameter of Mars can increase from only 3–4 arcsec at unfavourable times to 25 arcsec or more at the closest oppositions (see Figure 1); meanwhile, its brightness may increase by several magnitudes. These predictable yet irregular cycles of enhanced visibility provided a shared temporal framework within which astronomers, journalists and readers across Europe, North America and Australia engaged with Mars as a periodically intensified object of public attention.

Such oppositions occur roughly every twenty-six months, but only some are particularly favourable, depending on the relative positions of the Earth and Mars in their elliptical orbits. Mars follows a markedly eccentric path around the Sun (eccentricity ≈ 0.093), ranging from roughly 203–209 million km at perihelion to more than 240 million km at aphelion, whereas the Earth's orbit is comparatively circular: "Our orbit is very nearly a circle, the greatest variations in our distances from the sun being but 3½ million miles [5.6 million km]." (*Maffra Spectator*, 12 November 1894).

When opposition coincides with Mars being near its perihelion and the Earth near her aphelion, the distance between the two planets can fall to little more than 56 million km. At unfavourable oppositions, by contrast, Mars may remain nearly 80 million km away, while at superior conjunction the two planets are separated by more than 370 million km. These figures circulated widely in popular reporting, sometimes accurately and sometimes with conspicuous inflation (see also Section 5.4), underscoring how orbital mechanics became a central explanatory (and rhetorical) tool in Mars journalism (e.g., P.L.M., 1883; Jones, 1888; *Leader*, 27 August 1892; *Maffra Spectator*, 12 November 1894; *Australian Star*, 31 October 1896; and subsequent reprints).

Beginning with the particularly close opposition of 1877 (Earth–Mars separation ~ 56 million km), a sequence of favourable encounters coincided with improved telescopic capabilities—including the development of Yerkes Telescope in Chicago and the Great Lick Telescope in California—and the expanding reach of the popular press, producing sustained international interest in Mars. It was during these close approaches that observers believed the limits of telescopic resolution were being tested, which encouraged both cautious observation and speculative interpretation. Although popular reports sometimes exaggerated Mars's proximity to the Earth (see Section 5.4), the true significance of favourable oppositions lay not in extreme closeness but in the resulting increase in apparent angular size. Table 1 summarises the principal oppositions between 1877 and 1899 that shaped both observational practice—grounded in large refracting telescopes, silver-on-glass reflectors and micrometric measurement techniques—and press attention during this period. Exaggerated numerical claims regarding Mars's proximity, as commonly found in popular reporting, are discussed further in Section 5.4.

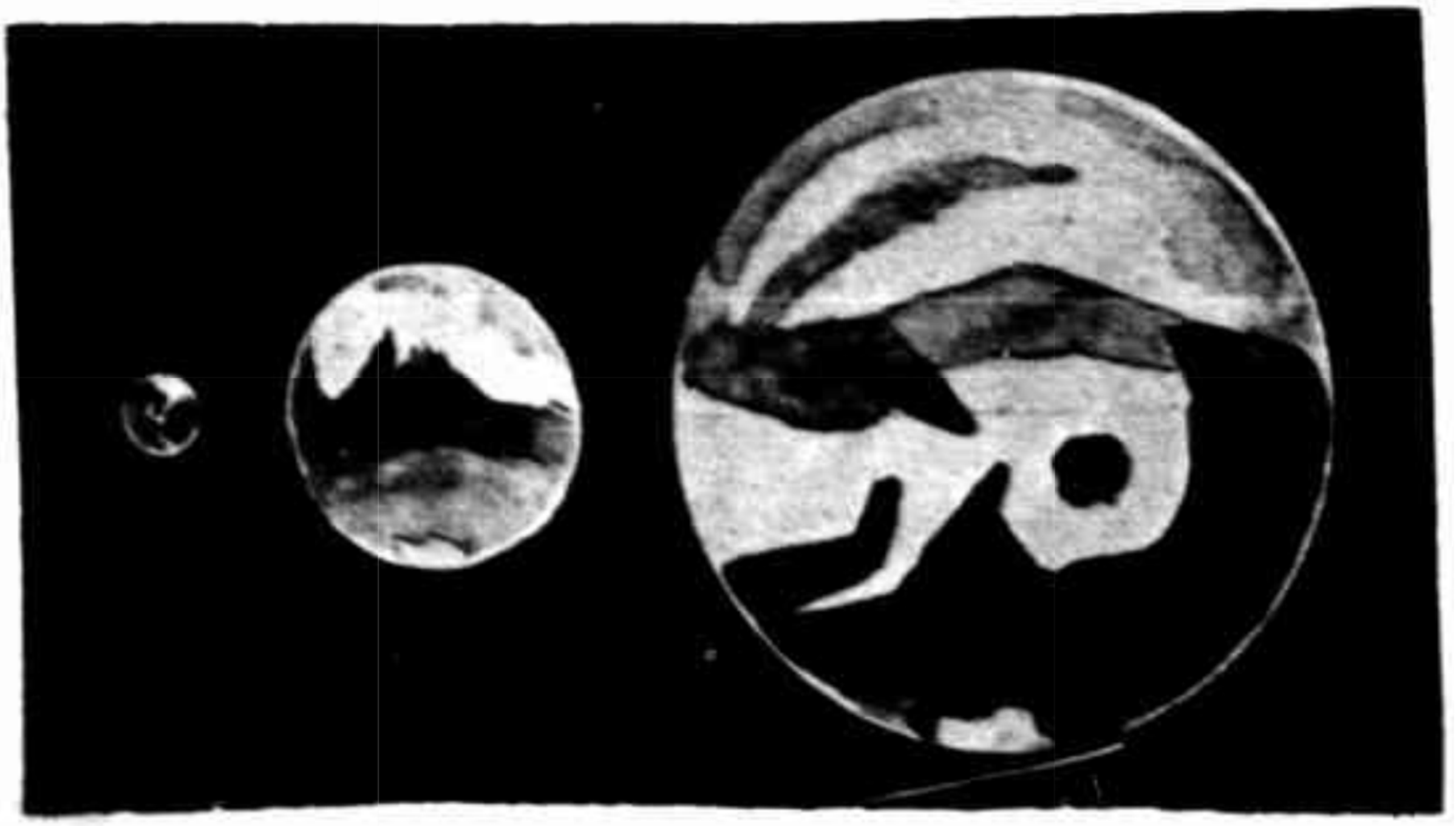

**Figure 1.** "Apparent dimensions of Mars at its extreme and mean distances." (*The Australasian*, 6 April 1895: p. 28; no known copyright restrictions.)

**Table 1**: Favourable Mars oppositions, 1877–1899.

| Opposition year | Closest Earth–Mars distance (million km) | Maximum apparent diameter (arcsec) | Observational significance |
|---|---|---|---|
| 1877 | ~56 | ~25 | Exceptionally favourable |
| 1881–1882 | ~78 | ~17 | Good |
| 1892 | ~56 | ~24–25 | Exceptionally favourable (particularly for southern observers) |
| 1894 | ~66 | ~21 | Very favourable |
| 1896 | ~78 | ~17 | Good |
| 1899 | ~95 | ~14 | Moderate |

Note: Distances are rounded to the nearest few million kilometres. Even at the most favourable oppositions, Mars does not approach Earth more closely than about 55–56 million km. The variation in apparent diameter reflects changes in the Earth–Mars distance driven by orbital geometry, including the distinction between perihelic and aphelic oppositions. The 1892 opposition was particularly favourable for southern observers owing to Mars' declination. Apparent diameters are approximate and vary slightly depending on ephemerides.

The historiography of this episode has tended to focus on metropolitan scientific centres,[1] major observatories and prominent figures such as Giovanni Virginio Schiaparelli (1835–1910; see Sheehan, 1996), Director of the Brera Observatory in Milan (Italy), and Percival Lowell (1855–1916; see, e.g., Crossley, 2011), Founding Director of the Lowell Observatory in Flagstaff, Arizona (USA). This emphasis has privileged metropolitan scientific institutions while largely overlooking how astronomical ideas were received, rearticulated and sometimes constrained within colonial media environments that nonetheless formed part of the same global information system. In colonial contexts, access to scientific information was mediated almost entirely through newspapers and the popular press. Indeed, by the late nineteenth century the Australian press was deeply embedded in global communication networks and served an increasingly literate and scientifically curious readership. As a result, Australian newspapers provide a valuable record of how planetary astronomy was encountered beyond the centres of discovery.

In this paper I examine how Mars was interpreted in the Australian press between 1875 and 1899, as part of a wider network of nineteenth-century scientific journalism linking Europe, North America and the colonial press. I adopt a comparative perspective, exploring how those interpretations were transmitted, negotiated and sometimes moderated within an Australian colonial press context. The focus is on the meanings attached to Mars in newspaper reporting: how observations were explained, how speculative claims were framed and how the possibility of life or even alien intelligence on Mars was presented to readers. In doing so, I explore the rhetorical and conceptual space in which scientific reporting eventually descended into imagination, analogy and debate.

Australian Mars reporting was largely derivative, relying heavily on overseas sources transmitted via telegraph or reprinted from foreign newspapers. Nevertheless, these items were not reproduced uncritically (see also Section 5.4 and Paper 2, the companion paper). Editorial choices, headlines, and patterns of repetition and syndication shaped how Mars was understood, while occasional local commentary and contextualisation reveal the ways in which global scientific narratives were absorbed into colonial culture. As such, the Australian press functioned not merely as a conduit for imported science but as a site of interpretation in its own right.

The period of interest ends in 1899 (see Section 2 for justification), on the eve of the twentieth century and approximately coinciding with the publication of Herbert George Wells's (1866–1946) *The War of the Worlds* (1898). By this point, many of the themes that would later dominate popular imaginings of Mars—canals, dying worlds and advanced but alien intelligences—were already firmly established in newspaper discourse. Examining this material allows us to recover a pre-fictional moment in which Mars was still framed in the popular press primarily as a scientific problem, albeit one rich in speculative possibility. By the turn of the twentieth century, the expansion of illustrated mass-circulation magazines, public lecturing circuits and speculative astronomical writing in Britain, France and the USA had begun to reshape both the authority and the tone of Mars discourse.

This study draws on a systematically compiled collection of Australian newspaper articles identified through keyword-based searches of the *Trove* digitised newspaper archive (https://trove.nla.gov.au/; see Section 2). The emphasis is mostly qualitative rather than quantitative: close reading is used to trace shifts in tone, emphasis and interpretive confidence over time. I pay particular attention to how scientific authority was invoked, how uncertainty was managed and how analogies to our Earthly experience—especially technological and environmental—were deployed to make Mars intelligible to readers.

Together with Paper 2, which analyses the circulation, attribution and temporal structure of Mars reporting in the Australian press, this study contributes to a more complete understanding of how planetary astronomy entered public discourse in late nineteenth-century Australia within a globally circulating astronomical news economy. Set against the better-documented Mars excitement in Europe and the USA (see the historiography cited above), where interpretive authority was often associated with prominent individuals and institutions, the present paper focuses on interpretation rather than transmission. I aim at highlighting the active role played by newspapers in shaping colonial engagements with one of the most evocative astronomical subjects of the period.

## 2. SOURCES, SCOPE AND METHODOLOGICAL APPROACH

This study is based on a systematically compiled corpus of Australian newspaper articles published between 1875 and 1899 and retrieved from the National Library of Australia's *Trove* digitised newspaper archive. The compilation was assembled through targeted keyword searches designed to capture substantive references to Mars as an astronomical object while minimising irrelevant results caused by the limitations of optical character recognition (OCR).

Search terms included |*planet Mars canals*|, |*Mars Schiaparelli*| and combinations of |*"planet Mars"*| with the names Hall, Pickering or Lowell. The inclusion of the term *planet* in two of the three systematic searches was found essential for filtering out spurious matches arising from homographs such as *mare* or abbreviated forms of *March*. While no keyword strategy can recover every relevant item, this approach produced a coherent and representative body of material reflecting what Australian readers were likely to encounter; see Paper 2 for a more detailed justification. Combined with manual sifting of relevant from non-relevant search hits and subsequent careful analysis and correction of the OCR-generated Trove entries, this approach resulted in a database containing 1040 articles.

Note that whereas this keyword strategy focuses on widely cited figures such as Schiaparelli and Lowell, other contributors to Mars debates—including Camille Flammarion (1842–1925) and, later, Eugène Antoniadi (1870–1944)—also appear in Australian newspaper coverage. My choice to focus on the search terms referenced above rather than attempt to include all contributors to these debates does not materially alter the recurring rhetorical patterns analysed here, which are anchored in thematic reporting practices rather than individual actors. Figures such as Flammarion appeared regularly in Anglophone newspaper discussions of Mars (e.g., Flammarion's name occurs 85 times in my database), although typically in more speculative or popularising contexts than those examined here. Their absence from the search terms used here reflects the thematic focus of the keyword search rather than their historical significance.

Once Mars had entered the public consciousness, which was strongly spurred by the discovery of the Martian satellites in 1877, Mars excitement in the Australian press followed the natural rhythm of major and minor oppositions. Significant increases in press coverage occurred in 1877, 1882, 1888, 1892 and 1894, whereas the baseline level gradually increased over the period of interest. Articles were ordered chronologically by first appearance, with subsequent reprints noted but not treated as independent sources. This structure allows interpretive shifts to be traced over time without artificially inflating the prominence of frequently reprinted material. The emphasis throughout is qualitative: close reading is used to examine language, metaphor, attribution and framing rather than to quantify coverage volume.

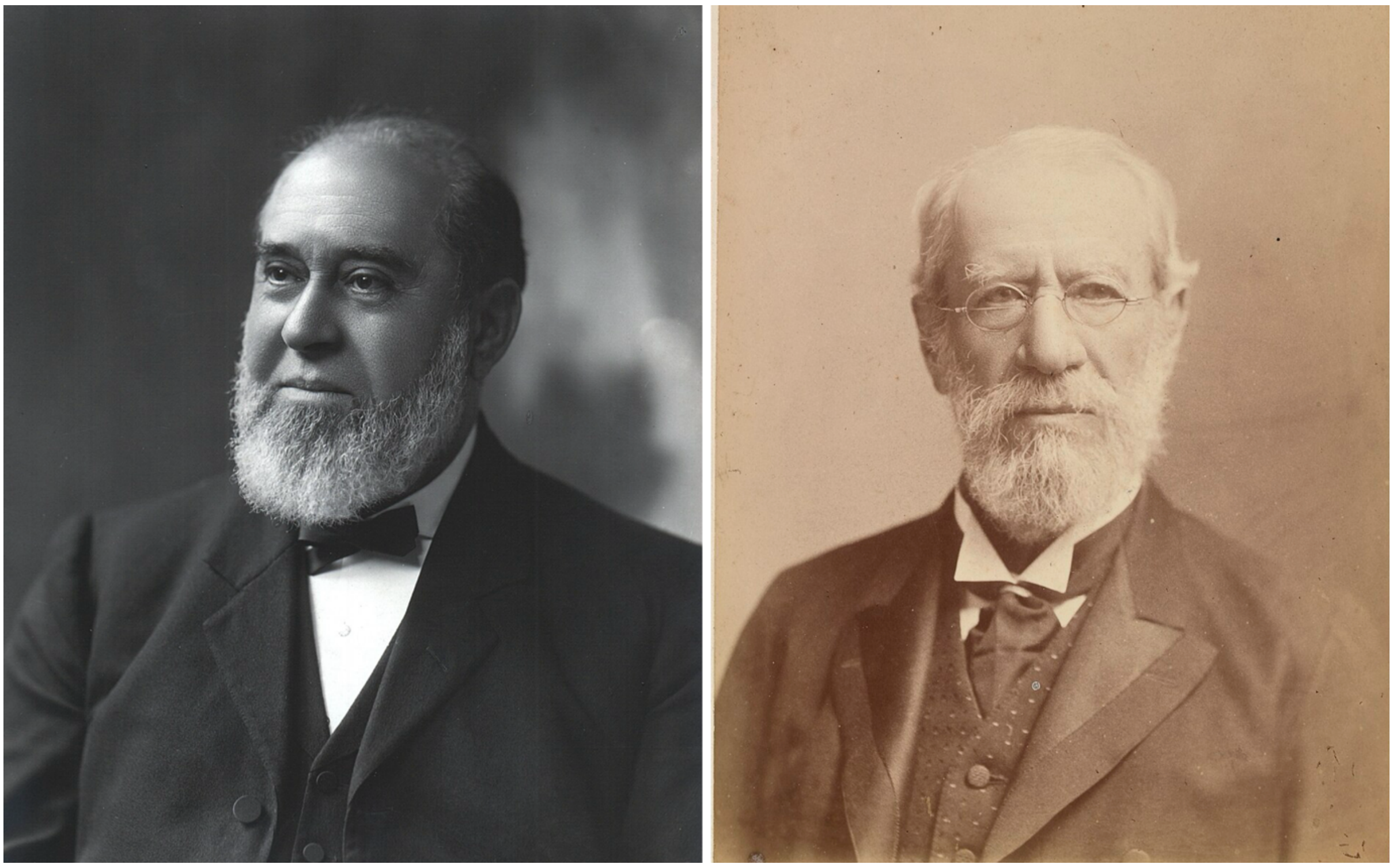

**Figure 2.** (*left*) Henry Chamberlain Russell, prior to 1907. Unknown photographer. (Smithsonian Institution Libraries/Dibner Library of the History of Science and Technology, Accession No. SIL 14-R004-06; no known copyright restrictions); (*right*) Robert J. L. Ellery, 1 January 1908. Photographer: T. Humphrey and Co. (State Library of Victoria, H24338; public domain).

The article collection consists predominantly of reprinted material originating primarily from British, French and American scientific centres transmitted via telegraph or syndication, supplemented by locally authored commentary contributed by a small but recurrent group of colonial 'savants'. These included (state) Government Astronomers such as Henry Chamberlain Russell (1836–1907; see Walsh, 1976; Bhathal, 1991) in New South Wales and Robert Lewis John Ellery (1827–1908; see Gascoigne, 1972) in Victoria (see Figure 2), scientifically literate correspondents and newspaper editors who provided explanatory framing or critical context. Such contributions were typically invoked at moments of heightened interest—especially around favourable oppositions—where they served to contextualise overseas claims rather than to advance original theories. These locally authored contributions were produced by individuals already embedded within international correspondence networks, observatory cultures and learned societies, even when they adopted a cautious or restrained public posture. Their institutional positions conferred a form of epistemic legitimacy that newspapers could mobilise selectively, particularly when moderating or contesting more speculative overseas claims.

These locally authored interventions were commonly positioned alongside syndicated material, lending credibility without displacing overseas authority. The authority of these local contributors derived less from discovery than from institutional position, observational experience or editorial credibility, and their interventions were generally received as moderating or interpretive rather than sensational, e.g.:

> The conflicting nature of the testimony renders it necessary that we should hesitate before accepting the evidence of the existence of the so-called canals as authentic, and still more that we should avoid the rashness of speculating on the possibility of executing great engineering works in Mars on account of the comparative smallness of the force of gravity on the surface of the planet. There is, however, no harm in such speculations, provided we are not called upon to have any faith in them. (Œdipus, 1882).

This reflects the realities of late nineteenth-century colonial journalism and is treated here not as a limitation but as a defining feature of the source base. The focus is therefore on interpretation within the press rather than on original scientific production.

The period of interest ends in 1899, a cut-off chosen for both historical and analytical reasons. By this point, the major nineteenth-century oppositions that had sustained popular interest in Mars—particularly those of 1877, 1892 and 1894—had passed, and the interpretive framework through which Mars was discussed in the press was largely settled. Key themes such as canals, planetary habitability and long-term decline had become familiar and formulaic, appearing with diminishing novelty in Australian newspapers.

Ending the study in 1899 also avoids conflating nineteenth-century scientific journalism with the changing cultural and media dynamics of the early twentieth century, including the institutionalisation of speculative astronomy, the expansion of mass-market illustration and the growing influence of science fiction,[2] thereby preserving the analytical coherence of nineteenth-century press-based astronomy. The cut-off thus preserves the coherence of a distinct phase in which Mars functioned primarily as a scientific and journalistic object rather than as a fully developed literary or popular-cultural icon. Although imaginative literature increasingly engaged with Mars after 1898, such works operated within a different discursive register from the scientific journalism examined here.

This does not imply a cessation of interest in Mars after 1899 (which may be a subject for further study) but rather marks the conclusion of a formative period in which the planet's meanings were first stabilised in the Australian press. Throughout this paper, emphasis is placed on understanding how Mars was rendered intelligible and meaningful to Australian readers. The methodological approach complements that of Paper 2, which focuses on circulation, attribution and temporal structure, allowing the two papers together to provide a comprehensive account of Mars in the Australian press in the late nineteenth century.

## 3. DISCOVERY, OBSERVATION AND THE LANGUAGE OF CERTAINTY

### 3.1 Mars as an observational object

In Australian newspaper reporting of the late nineteenth century, Mars initially appeared not as a speculative world but as an object of disciplined observation. Early articles gave explicit attention to the planet's orbital characteristics, noting the regularity of its oppositions, the marked variation in the Earth–Mars distance caused by orbital eccentricity and the resulting changes in apparent brightness and angular size (see Section 1 for quantitative details). Such discussions framed oppositions as privileged moments for observation, when Mars was rendered temporarily available for detailed scrutiny, e.g.:

> One of its nearest oppositions will take place on the 5th of September next [1877], on which occasion Mars will then appear sixty times larger than when at its greatest distance from the earth. (*Sydney Morning Herald*, 11 August 1877).

> In the autumn of the present year [1877] the planet Mars will present [an] appearance of unusual splendour; and as he will not be seen under such favourable conditions again during the present century, or indeed during the lifetime of any of the astronomers now living, considerable interest is attached to the circumstance, and preparations are being made in all observatories for the careful study of the planet's position and appearance. (*Dalby Herald and Western Queensland Advertiser*, 8 September 1877).

Reports frequently highlighted the role of improved instruments, particularly of large reflectors such as those at Melbourne and Washington with increased aperture and resolving power, as well as the establishment and upgrades of observatories equipped for planetary work. Together with references to favourable atmospheric and geographical conditions, specifically for Australian observers, these factors positioned Mars as a legitimate and increasingly accessible subject of scientific investigation rather than as an object of conjecture. This emphasis on careful measurement and favourable observing circumstances is illustrated particularly clearly by Australian press coverage of the 1877 opposition and the discovery of Mars's satellites (see Section 5.1), where local observatories were portrayed as active—if ultimately time-constrained—participants in an international observational campaign.

Australian newspaper reporting repeatedly emphasised the favourable visibility of Mars from southern latitudes, implicitly positioning Australia within the global geography of planetary observation, e.g., "… the planet Mars should be of special interest to Australians, inasmuch as opportunities for observing him are more favourable in the southern than in the northern hemisphere" (P.L.M., 1883). This reflected that, at favourable oppositions, Mars can reach declinations of around –20° to –30°, placing the planet high in the sky for southern observers. At such altitudes, atmospheric effects are reduced. At many northern observatories, by contrast, Mars would appear lower in the sky and the planet was therefore more affected by atmospheric distortion (see, e.g., Sheehan, 1996). This framing did not claim priority of discovery, but it did assert observational relevance: Australia appeared as a location from which overseas claims might be verified, refined or questioned. This position was further reinforced by reports that Australian Government astronomers received requests from their counterparts in Europe and, especially, the USA for observations during favourable oppositions, reflecting an international recognition of the importance of southern-hemisphere data. In this way, southern observations did not merely replicate northern findings but supplied temporal and geometrical perspectives unavailable to European and North American observers. Australian observations thus occupied a structurally necessary position within an international division of astronomical labour shaped by planetary geometry rather than by institutional hierarchy alone.

As a case in point, the English astronomer Richard Anthony Proctor (1837–1888; see Hutchins, 2004) appealed to his Australian counterparts for assistance: "Some very strange observations had been made by the Italian astronomer Schiaparelli, and he would like Australian astronomers to endeavour to verify these observations in the clear southern atmosphere they possessed" (*South Australian Register* and Adelaide's *Evening Journal*, 20 July 1880). Meanwhile, American observers, including William Henry Pickering (1858–1938, brother of Edward Charles Pickering, Director of the Harvard College Observatory; see Ashworth, 2023) emphasised, "It is very desirable that observations … in the longitude of Australia since certain features of the planet can be seen from there that are at the time obscured in other longitudes" (*Sydney Morning Herald*, 24 May 1894; and subsequent reprints). This positioning was reinforced by explicit requests from northern observatories for southern observations, a dynamic that newspapers reported as evidence of Australia's observational relevance within global astronomy. In this way, southern-hemisphere visibility functioned as a form of epistemic capital within the press, reinforcing the legitimacy of Australian engagement with Martian astronomy.

This observational framing served an important rhetorical function. By emphasising measurement, visibility and method, newspapers established a tone of authority and restraint. Mars was presented as knowable through proper scientific practice, and readers were invited to regard new reports as extensions of established astronomical inquiry rather than as conjecture or fantasy. Southern-hemisphere observational authority later found its most visible expression in the public interventions of Russell and Ellery (see Sections 5.2 and 5.3).

Following the favourable opposition of 1877, Schiaparelli reported the presence of linear markings (*canali*,[3] in the meaning of channels or grooves, not artificial canals) on the Martian surface (Schiaparelli, 1877–1878; 1895). This led to a subtle but significant shift in newspaper discourse. Initially described in cautious terms, these features were often introduced as 'streaks', 'lines' or 'channels', with explicit acknowledgment of the limitations imposed by distance, atmospheric conditions and optical resolution. Australian reports frequently attributed these observations to European astronomers, particularly Schiaparelli, whose work at Milan's Brera Observatory was regularly cited. In these early accounts, the emphasis remained firmly on observation rather than interpretation. The existence of the features was reported as a matter of record; their nature and origin were left open. This careful presentation reflects a broader journalistic strategy of maintaining scientific credibility while conveying potentially provocative information. Over time, however, the language used to describe Martian surface features began to shift. The term 'canals', once introduced, carried connotations that extended beyond neutral description. Although newspapers often noted the ambiguity of translation and interpretation, the word itself invited analogies to Earthly engineering and intentional design (see Section 4).

In Australian reporting, this transition is often subtle rather than explicit. Articles might combine careful observational statements with speculative possibilities, allowing implication to emerge without direct assertion. Phrases suggesting regularity, symmetry or a systematic approach appear alongside reminders of uncertainty, thereby creating tension between caution and curiosity. This balancing act enabled newspapers to explore the implications of the observations without abandoning the appearance of scientific restraint.

### 3.2 Managing uncertainty

Uncertainty is a persistent theme in Mars reporting during this period. Newspapers routinely acknowledged the difficulties of observation, the possibility of optical illusion and the provisional nature of current understanding, e.g., “A curious misapprehension respecting the hydrography of Mars has long been entertained, owing to the mistranslation of a word” (*The Age*, 21 January 1893). Such caveats functioned as safeguards against overinterpretation, reinforcing the authority of the reporting by demonstrating awareness of its limits. At the same time, uncertainty did not prevent the circulation of increasingly suggestive ideas. Instead, it provided a framework within which speculation could be entertained responsibly. By presenting multiple viewpoints or attributing interpretive claims to named astronomers, newspapers created space for debate while maintaining editorial distance.

Australian newspapers also positioned local observation as a means of testing overseas claims. Reports describing discoveries or interpretations made in Europe or North America were often followed by references to whether similar features had been observed from southern stations or Australian observatories. Even when such confirmation was inconclusive, the act of comparison reinforced Australia’s role as an observational participant rather than a passive recipient of scientific news. Mars thus became a subject through which global claims could be scrutinised locally, lending Australian reporting an evaluative dimension grounded in observation rather than theory. This approach allowed Australian readers to engage with the unfolding story of Mars without requiring them to resolve its ambiguities.

Throughout the initial phase of Australian Mars reporting, spanning roughly from the late 1870s to the early 1890s, scientific authority was grounded primarily in observational credibility. Reports frequently referenced observatories, instruments and scientific practice, with individual names serving as markers of expertise rather than as drivers of interpretation. This emphasis on institutional and methodological authority contrasts with later developments, in the mid-to-late 1890s, when Mars reporting increasingly became organised around prominent individuals whose interpretive claims came to dominate the narrative. In the earlier period, however, Mars remained framed as a shared scientific problem, open to investigation but resistant to definitive conclusions.

Between 1875 and the mid-1880s, Australian newspaper reporting established Mars as a scientifically respectable object of inquiry. The gradual introduction of linear features and the cautious adoption of the term ‘canals’ marked the beginning of a shift from purely descriptive astronomy towards interpretive possibility. This transition was managed through careful language, explicit acknowledgment of uncertainty and reliance on observational authority. These early patterns laid the groundwork for the more expansive and contested interpretations that would follow. In the next section, I will draw the reader’s attention to how the concept of canals was elaborated through analogy and engineering imagery and how these ideas resonated with broader nineteenth-century understandings of technological progress.

## 4. CANAL ENGINEERING AND PLANETARY ANALOGY

### 4.1 The canal as a nineteenth-century idea

By the late nineteenth century, the canal had become a powerful and widely understood symbol of large-scale technological capability (e.g., Headrick, 1981: Chaps. 4–5). Canals represented large-scale planning, sustained labour and the mastery of natural constraints through engineering. When Australian newspapers began to use the term ‘canals’ in relation to Mars, they therefore invoked a culturally mature concept. Even where journalists acknowledged that the term originated as a translation of *canali* and did not necessarily imply artificial construction, its repeated use carried interpretive weight. The canal was not merely a neutral descriptor; it was a familiar form of intentional infrastructure. Yet,

> Unfortunately canal to the public means a channel for ships cut out at great cost and labour, and therefore they thought there must be people and civilisation and commerce in Mars if there be canals, and that there were canals was vouched for by the astronomer's unfortunate translation. This has misled tens of thousands of persons. (*Sydney Morning Herald*, 10 December 1892).

Australian newspaper accounts frequently explored the implications of Martian canals through analogy with terrestrial engineering projects. Descriptions of linear features as ‘systems’ or ‘networks’, sometimes still under construction (e.g., *Maitland Mercury and Hunter River General Advertiser*, 5 July 1888; and subsequent reprints), encouraged readers to imagine Mars as a world shaped by rational planning. The regularity and apparent global scale of the features further reinforced this impression. Such analogies did not always assert artificiality directly. Instead, they operated by accumulation: repeated references to order, scale and connection gradually framed the canals as something more than natural formations. In this way, newspapers could advance provocative interpretations while maintaining a semblance of scientific caution.

The period of interest coincided with widespread public awareness of major canal projects on Earth, most notably the Suez Canal and, from the 1880s onwards, renewed efforts to construct a canal across the Isthmus of Panama. (The Panama Canal was eventually opened in 1914, after the USA took over the initially unsuccessful attempts by French engineers in 1904.) References to terrestrial canal projects framed Martian speculation within a global discourse of industrial modernity, engineering ambition and environmental mastery. Australian newspapers repeatedly invoked these terrestrial engineering works when reporting on the so-called canals of Mars, sometimes explicitly and sometimes through irony or analogy—"To many people the Panama Canal seems an enterprise as dim and as much out of human reach as the canals which astronomers report in the planet Mars" (*Shepparton Advertiser*, 19 February 1889). "Mars is a place, in fact, where all good Panama promoters might hope to go" (*Maitland Daily Mercury*, 18 October 1898; citing London's *Pall Mall Gazette*).

In doing so, they situated Martian features within a familiar vocabulary of global engineering. The Suez Canal was routinely described as a "wonder of the world" (e.g., *Australische Zeitung*, 3 October 1882), a benchmark against which the scale and ambition of Martian canals could be measured. Reports noted that if features on Mars were truly artificial, they would necessarily eclipse the Suez Canal in magnitude (e.g., *Geelong Advertiser*, 13 March 1897), rendering even the most celebrated terrestrial works comparatively insignificant "… and even eclipse the one at Sale [Victoria]" (e.g., *Maffra Spectator*, 12 November 1894); "… for a complicated scheme of irrigation extended over the entire area of a planet is a bigger problem than any terrestrial engineer would care to take in hand; compared with it, such pieces of work as the Forth Bridge [at Edinburgh, Scotland] or the Suez Canal are almost infinitely insignificant" (*Riverine Herald*, 26 October 1896).

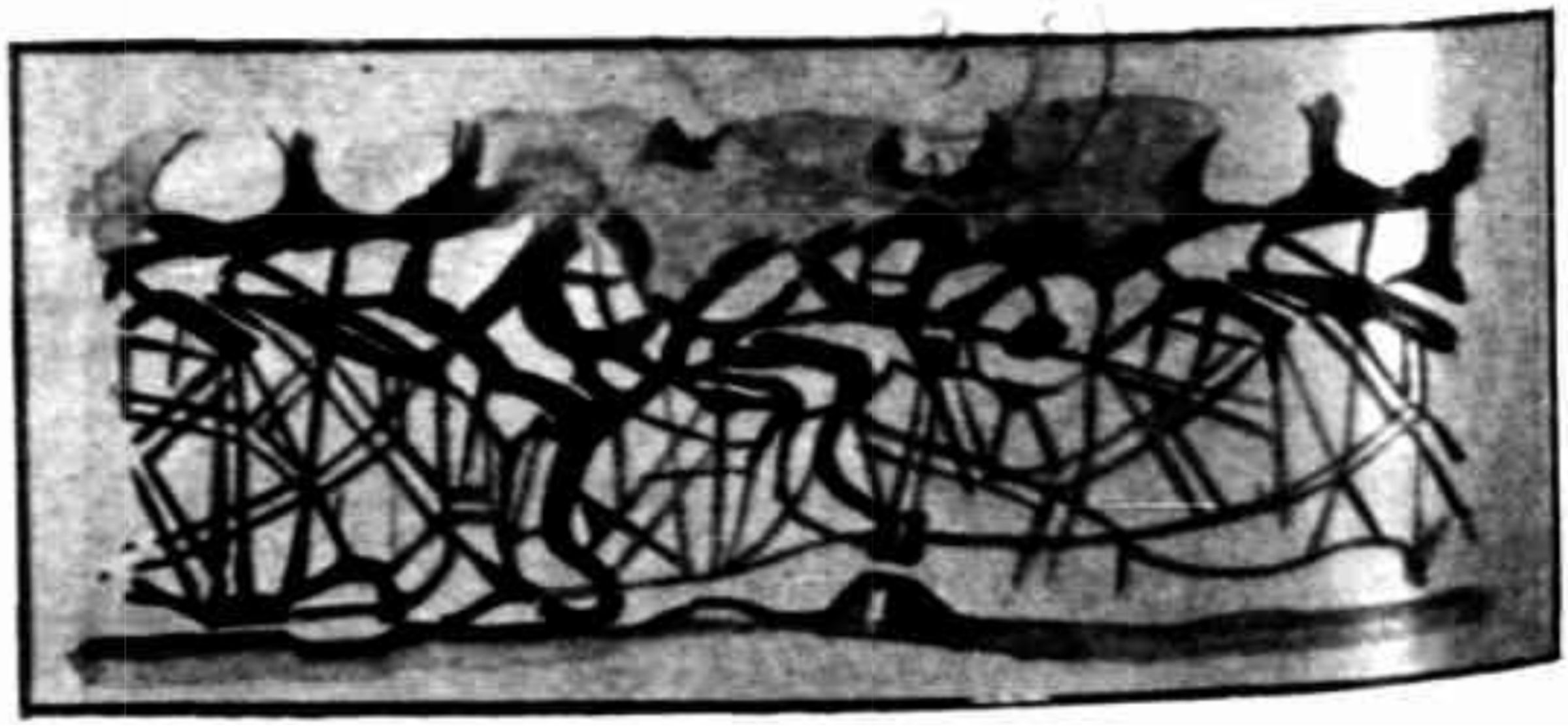

**Figure 3.** "The 'channels' of Mars as recorded by Schiaparelli after their doubling." (*The Australasian*, 6 April 1895: p. 28; no known copyright restrictions.)

Such comparisons were not confined to sober exposition. Satirical and speculative articles alike drew on the language of contemporary canal building to animate Martian narratives:

> As if that of Panama was not sufficient, opinion is occupied with the presumed canals in the planet Mars. … One canal, communicating with a vast fresh water ocean, has a lock. Possibly it was that circumstance which caused M[onsieur]. [Ferdinand Marie] de Lesseps [1805–1904; French developer of the Panama Canal] to change his mind and apply the wrinkle to his Pacifico–Atlantic ditch. According to the latest astronomical interviewer, Mars would seem to resemble not a little Holland;[4] not only by its canal system, but by its windmills, barge traffic, and winged inhabitants. Our planet can at least boast of one "Flying Dutchman." Many speculators would like to know what it cost to construct a ship canal in Mars, how many times the capital was augmented before the opening day, if a lottery loan had been resorted to, and what recompense was accorded to the piercer. … But we must wait till the Eiffel Tower be completed [15 May 1889] to have an Asmodean [supernatural] peep into the mansions of our planet neighbours. (*Ballarat Star*, 15 August 1888).

Australian newspapers imagined Martians as having "duplicated their Suez Canal" (see, e.g., Figure 3) or completed a polar "Panama Canal" (e.g., *Argus*, 20 August 1888; *The Age*, 1 December 1894; *Australian Town and Country Journal*, 24 October 1896), while Earth's engineers were still mired in financial, political and technical delay:

> … while the lottery loan that is going to put the finishing touch on the enterprise of M[onsieur]. de Lesseps has not yet been raised and now they [the Martians] are supposed to have carried out an irrigation work by the side of which the Mildura experiment[5] and all the Chief Secretary's pet schemes in addition look very small indeed. From this it would seem that the Martian must be, man for man, a considerably finer fellow, than the European, or even the Australian. (*Ibid.*)

Elsewhere, writers calculated that the total length of the Martian canal system—sometimes estimated at tens of thousands of miles—would be equivalent to hundreds of thousands or even millions of Suez Canals (e.g., *Australian Star*, 27 April 1895), a move that simultaneously conveyed scale and strained credibility. These numerical analogies often served a critical function, undermining claims of artificial origin by pushing them to the point of absurdity.

At the same time, the canal analogy was readily mobilised in earnest discussions of Martian habitability. Drawing on contemporary reporting about irrigation schemes, river management and interoceanic transit, newspapers framed Martian canals as responses to environmental necessity, particularly the scarcity of water on an aging planet. In this context, Mars was imagined as a world whose inhabitants, like modern engineers on Earth, had reshaped their environment through coordinated, planet-spanning works. That Australian newspapers could invoke Suez, Panama, Mildura irrigation or even local roadworks without extended explanation underscores how deeply embedded the canal concept had become in late-nineteenth-century public consciousness. The absence of sustained technical comparison does not diminish the significance of these resonances; rather, it reveals how the idea of canals—human, artificial and transformative—could be transferred almost effortlessly from Earth to Mars.

### 4.2 Water, climate and planetary management

Canals also carried implicit associations with water management and environmental control. Australian reports often linked Martian canals to theories of a drying planet, in which dwindling water resources required large-scale redistribution (e.g., Gibier, 1892; *Australian Star*, 31 October and 27 November 1896; among numerous other articles). This framing aligned with broader scientific discussions of planetary ageing and climatic decline. In this context, canals became symbols not only of intelligence but of adaptation. They suggested a civilisation responding creatively to environmental constraint, a theme that would recur in later interpretations of Mars as an older, colder world (e.g., *Ballarat Star*, 31 July 1886; Jones, 1888; and numerous other articles). Such narratives positioned Mars as both technologically advanced and environmentally precarious, inviting reflection on the relationship between progress and survival.

Although most Mars reporting drew on overseas sources, Australian readers encountered these ideas within a local environmental context shaped by aridity, irrigation and settlement. The notion of managing scarce water through engineered channels would have been particularly intelligible in a society grappling with drought and agricultural expansion. While newspapers rarely made this connection explicit, the resonance is nevertheless significant. Mars could be understood through experiences familiar to Australian readers, lending immediacy and plausibility to otherwise distant astronomical speculation.

By the late 1880s and early 1890s, the cumulative effect of engineering analogy was to establish an expectation of artificiality. Even cautious reports began to treat the possibility of intelligent design as a reasonable hypothesis rather than a remote conjecture. The canal concept thus functioned as a bridge between observation and speculation. This transition did not occur abruptly but through gradual rhetorical reinforcement. Each new report reaffirmed earlier associations, embedding the idea of canals within a broader framework of technological meaning. The adoption of the canal analogy marked a turning point in the Australian press's interpretation of Mars. By drawing on familiar concepts of engineering, water management and environmental adaptation, newspapers transformed ambiguous observations into intelligible and compelling narratives. These narratives did not merely describe Mars; they imagined it as a world shaped by intelligence and necessity.

In the next section, I will turn the reader's attention to the role of individual scientific authorities in addressing these interpretations. Named personalities increasingly came to dominate Mars reporting in the Australian press.

## 5. AUTHORITY, PERSONALITY AND INTERPRETIVE ESCALATION

As Mars reporting evolved in the Australian press, the construction of scientific authority underwent a noticeable transformation. Early articles relied primarily on institutional credibility, invoking observatories, instruments and common scientific practice. Over time, however, individual astronomers increasingly became focal points around which interpretation was organised. This shift reflects broader changes in late-nineteenth-century science journalism, in which named experts provided both explanatory clarity and narrative coherence. In the case of Mars, the growing prominence of individual scientists allowed newspapers to attach speculative claims to recognisable authorities, thereby legitimising interpretation without assuming editorial responsibility.

Although much of the authority cited in Australian Mars reporting derived from overseas scientists, the episode surrounding the discovery of Mars's satellites in 1877 (see the next section) demonstrates that Australian observatories were not merely passive recipients of expert opinion; they were explicitly positioned as contributors whose observations were anticipated within global astronomical networks.

### 5.1 Asaph Hall and the Discovery of the Martian Moons

Asaph Hall (1829–1907; see Hockey et al., 2007) entered Australian Mars reporting chiefly through his discovery, on 11–18 August 1877, of the Martian satellites, Phobos (fear) and Deimos (terror), with the U.S. Naval Observatory's 26-inch telescope during the 1877 opposition (Sheehan, 1996: Chap. 5). Although identified at the U. S. Naval Observatory, the discovery was rapidly evaluated, discussed and contextualised through observatories in Paris, Cambridge (Massachusetts, USA), Melbourne and Sydney. Press accounts of this achievement emphasised methodological rigour sustained through prolonged observation: e.g., "It was still thought best to wait for another look before formally announcing the discovery …" and "He is an able and learned mathematician, and an unostentatious and conscientious observer …" (Henry, 1877; and subsequent reprints). Hall's work was consistently framed as a triumph of sustained observation rather than interpretation—"… one of the most remarkable additions to modern astronomy …" (*Ibid.*)—reinforcing a view of Mars as a planetary system whose properties could be revealed through careful measurement at moments of unusual proximity:

> Mars has two satellites, the discovery of which is another of the triumphs of modern skill and mechanical art in perfecting our instruments. They were discovered … by means of the most powerful telescope at that time constructed. The discovery was not made by chance, but was the result of a systematic search. (Taylor, 1884).

Australian coverage of the discovery, however, did more than merely relay overseas news. Several newspapers explicitly situated Australian observatories within the same observational campaign, stressing both their instrumental capability and their geographical advantage. Reports noted that Sir George Biddell Airy (1801–1892), Great Britain's seventh Astronomer Royal (1835–1881), had requested that a close watch be kept at the Melbourne Observatory for "suspected satellites of Mars" (*Argus*, 16 October 1877: p. 4; and subsequent reprints) while similar efforts were undertaken at Sydney and Adelaide. Particular emphasis was placed on the Melbourne reflector, which was described as comparable in size and quality to the large reflector employed at the Paris Observatory, where confirmatory observations were also reported. Australia's southern latitude was repeatedly invoked as providing superior observing conditions, in contrast to Europe, where Mars appeared low in the sky during the critical period.

The failure of Australian astronomers to detect the satellites was therefore not presented as a deficiency of skill or equipment but because of delayed communication. Several articles expressed regret that news of Hall's discovery had not reached Australia earlier, when Mars was closer to the Earth and observational conditions more favourable:

> We further learn that information has reached the Melbourne Observatory to the effect that one of the presumed satellites was observed by M[onsieur]. [Paul] Henry [1848–1905][6] at the Paris Observatory on the night of August 27 [1877]. ... The telegram of Sir George Airy, requesting that a watch might be kept at the Melbourne Observatory for "suspected satellites of Mars" was received on the 24th September, and it was not until the 11th inst[ant] that the message from Washington arrived [at Melbourne]. It is to be

> regretted that communication was not made more promptly, for at the time of the discovery at Washington, and the confirmatory observations at Paris, the planet was much nearer to the [E]arth than he has since been. (*Argus*, 16 October 1877, p. 4; and subsequent reprints).

By the time formal notification arrived, the planet had already receded, thus limiting the prospects for independent confirmation. In this framing, Australian astronomers appeared as willing and competent participants in an international observational enterprise whose success depended crucially on timing rather than expertise:

> Our [Melbourne] astronomers … have done their best. An assiduous watch has been kept, with relays of observers, but not a trace of either satellite has been seen. Mr. H. C. Russell, of the Sydney Observatory, has been equally unsuccessful, as we learn from a letter which he has addressed to the *Sydney Morning Herald*. (*Ibid*.)

> Mr. Russel[l], the Government astronomer, has received a communication from Sir George Airey [*sic*], by which it appears that to Professor Hall, of the United States, is due the discovery of two satellites of the planet Mars, which has lately paid the [E]arth a very close visit. ... Mr. Russell has, he says, looked very carefully, on every available opportunity since he received the notification from Sir George Airey, but has not been able to discover either of them. (*Rockhampton Bulletin*, 30 October 1877).

> Sir George Airy, the Royal Astronomer of England, … particularly requested Mr. Ellery of the Melbourne Observatory … to make use of this time and to direct his observations especially toward determining whether Mars has one or more moons. … Mr. Ellery therefore began his observations with great zeal, as did Mr. Russell in Sydney and Mr. [Charles] Todd [1826–1910] in Adelaide. Yet fortune did not favour them, for on 16 August [1877] it appears that Professor Hall … discovered the suspected new celestial body—the moon of Mars—and, if he was not mistaken, even a second one. … During this observation, Mr. Todd noticed a second, smaller star even closer to Mars, though it appeared so faint that nothing certain could yet be decided; nevertheless, the suspicion is strong that it represents a second moon. (*Australische Zeitung*, 30 October 1877; own translation from German).

Importantly, Hall was rarely associated in the Australian press with speculation about canals or Martian life. His authority functioned instead as an anchor of restraint, embodying a model of planetary astronomy grounded in disciplined observation and empirical caution. When referenced in later discussions (e.g., Taylor, 1884), Hall's name served to underscore the credibility of Martian astronomy without advancing controversial interpretive claims. In this respect, the authority represented by Hall found its closest Australian parallels in the work of institutional astronomers such as Ellery and Russell, whose engagement with Mars was similarly framed as careful, methodical and sceptical of premature speculation.

The satellite episode thus provides an instructive contrast with later debates on Martian canals: where the moons demanded rapid, time-critical observation, the canals invited prolonged interpretation, allowing Australian newspapers—and, occasionally, Australian observers—to engage more fully with questions of meaning rather than detection.

### 5.2 Russell and the Management of Martian Credibility

Where Hall represented disciplined discovery at a distance, Russell embodied scientific authority exercised locally and in public view. From the early 1890s, Russell appeared repeatedly in Australian newspaper coverage as an active interpreter of Martian claims (e.g., see Figure 4), addressing both popular curiosity and professional controversy. His interventions were not confined to brief statements but included public lectures, learned society addresses, interviews and detailed explanatory essays, positioning him as the principal Australian mediator between overseas astronomical debate and colonial audiences.

Russell's responses consistently emphasised physical plausibility, instrumental limits and atmospheric physics:

> Mars observers think they see clouds. I have failed to do so, and in cases where something has temporarily covered well-known features of the planet, I think it has been haze more than cloud. … it would be impossible for clouds that we should call heavy ones to float in it. For particles of water to float there they must be very much more minute than these of which we know. (*Sydney Morning Herald*, 10 December 1892; and subsequent reprints).

During the 1892 opposition, when popular excitement over alleged Martian signalling reached a peak, Russell publicly rejected the notion that observed fluctuations in brightness represented intentional communication, attributing them instead to transient cloud cover and atmospheric effects (e.g., *Sydney Morning Herald*, 24 August 1894):

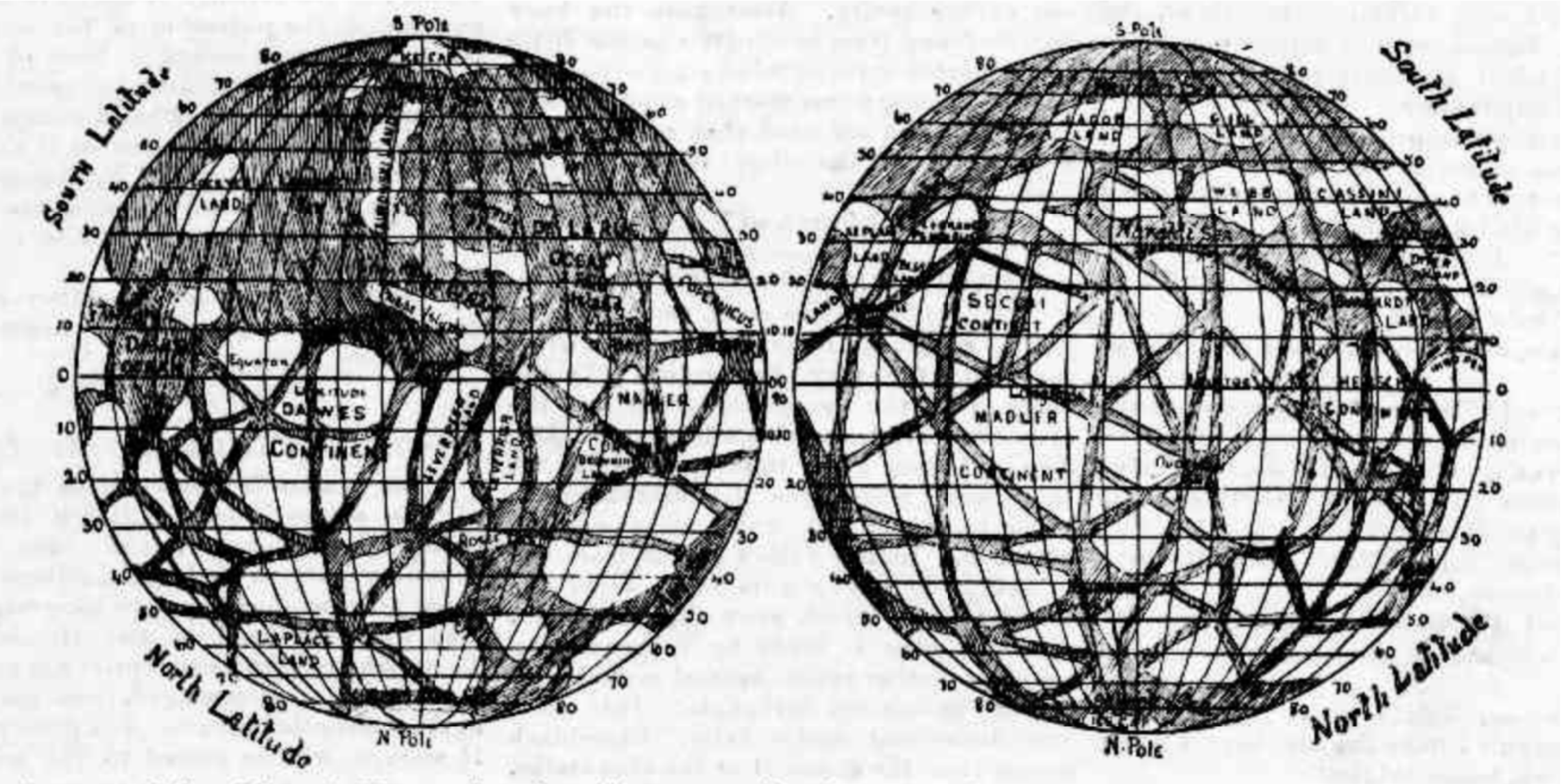

**A map of the Planet Mars, showing the recently discovered (supposed) continents and canals.**

**Figure 4.** "By the courtesy of Mr. H. C. Russell, the Government Astronomer, we are enabled to present to our readers a copy of the most complete and reliable map of the planet Mars available in the present stage of astronomical science." (*Illustrated Sydney News*, 18 February 1893: p. 18; no known copyright restrictions.)

> Considerable excitement was occasioned in the city [Sydney] and suburbs last night (says the *S[ydney]. D[aily]. Telegraph* of Friday last [12 August 1892]) by a curious phenomenon, which was to be observed in connection with the planet Mars. This usually bright star [*sic*], while shining with its ordinary brilliancy, would suddenly fade away, so as to be almost invisible, and then disappear just as suddenly and just as brightly as before. At times the change was by no means so sudden; though, strange to say, the periods during which the full light of the star was seen were strikingly regular, generally at intervals of about two minutes. Groups of people gathered at the street corners or stood on the kerbstones and gazed steadfastly at the planet, and having seen the change take place hurried to tell some one else of the new wonder. The result was that at half-past 8 nearly everyone in the city had been informed, and the number that shaded their eyes and gazed into space had increased considerably. Various suggestions as to the cause of the variation of the light were made, many being fully convinced that the inhabitants of Mars had adopted a system of signalling to the people of this earth. Mr. H. C. Russell, the [New South Wales] Government Astronomer, said he had not witnessed the eccentric behaviour of the planet in question, but that several inquiries in regard to the matter had been made. The occurrence was not at all uncommon, he said, and was caused merely by the passing of clouds, he was not inclined to regard seriously the suggestion that the inhabitants were endeavouring to "signalise." (*Brisbane Courier* and *Queensland Times, Ipswich Herald and General Advertiser*, 16 August 1892).

In responding with restraint, Russell framed himself not as a sceptic hostile to speculation but as a custodian of methodological restraint, repeatedly reminding readers that extraordinary claims demanded extraordinary observational evidence.

His December 1892 address to the Royal Society of New South Wales, widely reprinted across the Australian press (e.g., *Sydney Morning Herald*, 10 December 1892; and subsequent reprints), marked the most sustained Australian engagement with the question of Martian habitability. Russell did not dismiss the possibility outright; instead, he reconstructed it from first principles: "Sir Robert Ball [1840–1913; Royal Astronomer of Ireland] in a recent publication, has well said, 'the idea of signalling to Mars is a preposterous one.' Let us see why" (*Ibid.*). He then proceeded by discussing planetary mass, specific surface gravity, atmospheric retention and thermal balance, all in comparison with those properties on Earth:

> But I fear that the circumstances which, to so large an extent, make or mar us here, would not let us do very much on Mars, for the same freedom from the major parts of gravity, as we know it here, would enable the air to expand until our muscles, relieved of the outward pressure which our atmosphere exerts upon us, would relax, and we should he gasping for breath, and find the blood oozing out of the pores of

> our skin, because we should be in such an attenuated atmosphere as we should find upon the [E]arth at a height of about 10 miles [16 km]. (*Ibid.*)

By quantifying these constraints and invoking spectroscopic evidence from William Huggins (1824–1910; Huggins, 1867),[7] Russell shifted discussion away from imaginative analogy and towards comparative planetary physics:

> Many years ago Dr. Huggins examined the spectrum of the planet and found certain spectrum lines which did not belong to the solar spectrum. … the discovery by Dr. Huggins, one of the most able observers with the spectroscope in the world of these different lines in Mars, showed that they must be introduced by something in the atmosphere of the planet itself. (*Australian Star*, 24 August 1894).

This mode of argument positioned Australian astronomy as analytically mature rather than derivative.

Russell also occupied a more collaborative observational role. Correspondence from Pickering, Professor of Astronomy and Geodesy at Harvard University, explicitly requested Australian observations of specific Martian surface features during the 1894 observing season, citing Australia's advantageous longitude:

> In a letter to me dated 29th March Professor W. H. Pickering of Harvard College Observatory, says "It is very desirable that observations of the surface details of Mars should be made in the latter part of May and early in June in the longitude of Australia since certain features of the planet can be seen from there that are at the time obscured in other longitudes." … But Mars is only 10 seconds of arc in diameter, and in order to see those small details one must be able to see clearly a mark not more than one fiftieth of the planet's diameter. Telescopes must therefore be first-class, and probably less than 6in [15.2 cm] object glass will be insufficient. A number of your readers have the requisite telescopes. I hope they may be induced to help in the work. (*Sydney Morning Herald*, 24 May 1894; and subsequent reprints; see also *Australian Star*, *Sydney Morning Herald* and *Daily Telegraph*, 24 August 1894).

Russell's public appeal to readers with access to suitable telescopes to assist in this work demonstrates that Australian participation was not limited to institutional observatories but extended, at least rhetorically, to a broader community of competent observers. Although Russell himself reported difficulty in obtaining conditions sufficient for detecting fine surface detail—"In the [New South Wales] colony, so far as I am aware, no one has seen anything of these markings." (*Australian Star*, 24 August 1894)—the episode illustrates Australia's recognised place within international observing networks.

Finally, Russell emerged as a prominent critic of Lowell's claims. While careful not to issue categorical denunciations, he repeatedly expressed doubt that small instruments could reveal features invisible to the world's largest telescopes: "… he said it appeared to him that they claimed to have seen a great deal more with a small telescope than with the largest telescope in the world. 'It is,' said Mr. Russell, 'hard to believe that they have seen so much'" (*Australian Star*, 24 August 1894). In this respect, Russell functioned as a stabilising authority within Australian Mars reporting, consistently tempering enthusiasm with appeals to optical limits, repeatability and observational consensus.

Notably, Russell's relationship with the press was not always harmonious. A Sydney *Evening News* report of 13 September 1892 remarked pointedly, "What is being done and has been done in this respect at the Sydney Observatory we are not at the moment in a position to state …", adding that Russell "… often regards his observations of the stars as strictly private and confidential." Although framed with a degree of journalistic irritation, this comment provides rare insight into the professional norms governing late-nineteenth-century astronomical practice in Australia.

Russell's apparent reticence reflected a deeply ingrained observational ethic rather than institutional obstruction. Observations were commonly regarded as provisional until fully reduced, compared and interpreted, particularly in cases—such as Mars—where atmospheric conditions, optical limitations and observer expectation could strongly influence perception. From this perspective, withholding preliminary results was a mark of scientific responsibility, not evasiveness. Nevertheless, such restraint conflicted with the press's desire for immediate, authoritative commentary during moments of heightened public interest.

The *Evening News*'s sardonic comparison—suggesting that confidentiality might be appropriate for stars "… such as Cassiopeia …"[8] but not for Mars—reveals an important shift in expectations. Mars, unlike the fixed stars, had become a public object: its proximity, alleged canals and potential habitability rendered it a matter of popular concern rather than purely technical inquiry. Russell's insistence on observational reserve (e.g., *Sydney Morning Herald*, 24 August 1894) therefore placed him at odds with the evolving role of astronomy as a subject of mass communication. As a second example, reading between the lines in the *Australian Star* of 24 August 1894 (see also the *Clarence and Richmond Examiner* of 28 August), we encounter some additional journalistic frustration:

"He is not prepared to controvert the views of the Lick Observatory in the absence of information of what discoveries they have made which has caused them to declare that the weight of evidence is now in favor [*sic*] of no atmosphere."

This tension is significant for understanding Australian engagement with the Martian controversy. Russell did not refuse public participation; indeed, he later spoke extensively and publicly on Martian conditions. Rather, he sought to control the timing and framing of disclosure. In doing so, he implicitly resisted the transformation of astronomical observation into a form of real-time spectacle, a transformation that would soon be embraced by figures such as Lowell. Russell's stance thus represents a transitional moment in the relationship between scientific authority and public expectation within Australian astronomy.

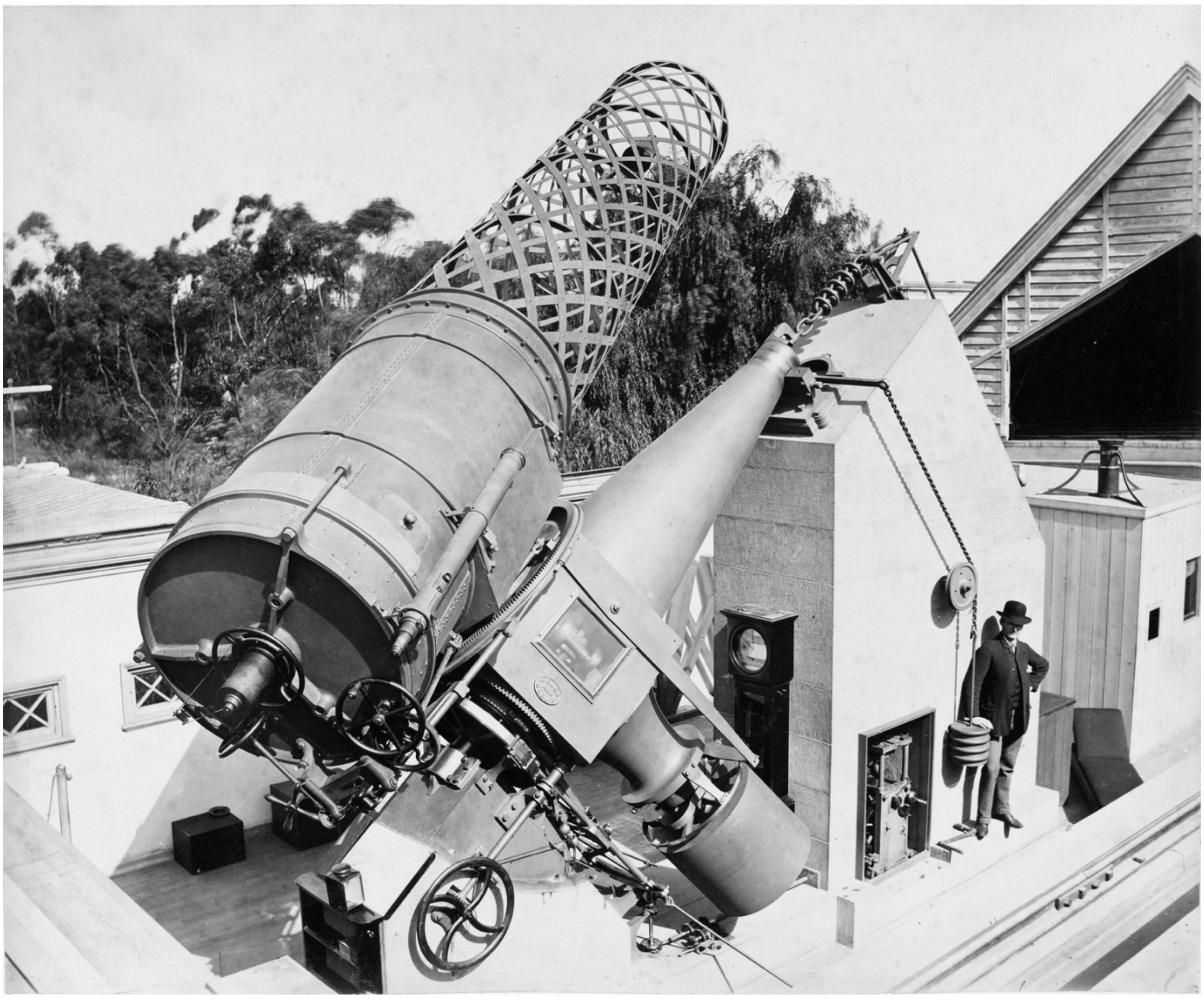

**Figure 5.** The Great Melbourne Telescope, a 48-inch (122 cm) equatorial reflecting telescope, at Melbourne Observatory, ca. 1875. (Museums Victoria; public domain.)

### 5.3 Ellery and the Limits of Interpretation

Ellery appeared even more frequently than Russell in Australian newspaper discussions of Mars, particularly during periods of heightened speculative reporting. Ellery's authority derived not only from his institutional position but also from the international reputation of the Melbourne Observatory and its Great Melbourne Telescope (see Figure 5), which was repeatedly invoked in press accounts as among the finest instruments in the southern hemisphere.

Ellery's public statements consistently resisted sensational interpretations of Martian surface features:

> If he did no more than relieve the popular mind from the uneasiness that was gradually taking possession of it on learning that the people in Mars were constructing straight canals hundreds of miles wide, he would still have done well … The desire on the part of pseudo-scientists to circulate wild conjectures among the masses whenever a heavenly body rises a little redder or sets a little paler than usual, cannot be too severely reprobated. (*Sydney Morning Herald*, 1 September 1888).

As early as the late 1880s, he openly criticised what he characterised as speculative French interpretations of Martian changes, including the idea that linear markings represented irrigation canals (e.g., *The Age*, 20 September 1888; *Argus*, 21 September 1888; *Daily Telegraph*, 30 August 1888; *South Australian Register*, 13 September 1888; *Argus*, 15 March 1889; and subsequent reprints). These objections were framed not as dismissals of observation, but as cautions against interpretive overreach: "There could be no doubt about the appearances, but there was about the interpretation of them. Astronomers, he thought, were waiting further developments, for there were inseparable difficulties in accepting explanations already ventured upon" (*Argus*, 15 March 1889; and subsequent reprints). In fact,

> French views upon the changes in the planet Mars, their cause and operation, were condemned as speculative, and the theory that the markings called "canals" were really used for irrigation purposes was ridiculed. (*Daily Telegraph*, 30 August 1888).

Ellery repeatedly emphasised the persistence of many Martian markings across oppositions, arguing that their stability was more consistent with optical, atmospheric or geological explanations than with large-scale artificial construction:

> It was more than probable that all the appearances would eventually be attributed to diffraction or optical effects, from conformation of vapours surrounding the planet, rather than of any objective change in its surface. (*Argus*, 15 March 1889; and subsequent reprints).

> According to Mr. Ellery, the real peculiarity of Mars is that he has two small moons that keep close to him all through his tedious journey, and swing round him with bewilderingly rapidity. It is surely improbable that people who lived in daily fear of being crushed by bodies so perilously close, would devote their time to cutting long street [*sic*] canals. (*Sydney Morning Herald*, 1 September 1888).

During the 1892 opposition, Ellery emerged as a vocal critic of telegraphed claims attributed to Pickering's Peruvian observations. He publicly doubted that discoveries made with Pickering's comparatively modest instruments in Arequipa, Peru, could exceed what had been achieved with superior telescopes in the northern hemisphere. Specifically, the *Argus* of 5 September 1892 pointed out such limitations in the context of Pickering's newly published observational details—eleven (or even up to 40) lakes and two mountain ranges—from Peru: "Professor Pickering had no instruments there of a large size, and those he had there were for spectroscopic purposes and not such as would be likely to lead to any new discoveries on the features of Mars."

> Mr. ELLERY points out that the professor is not likely to have made discoveries with inferior instruments in Peru that he has failed to make with the best instruments at Harvard, and attributes the concise circumstantiality of the telegram to the imagination of the newspaper correspondent. (*Argus*, 10 September 1892).

Australian newspaper editors were similarly dismissive of Pickering's claims, as exemplified in the *Argus* of 10 September 1892 and the *Gippsland Daily News* of 12 September 1892, for instance:

> There is a strong idea that [Mars] must be inhabited. But nobody has yet seen the Eiffel towers or Great Eastern steamers [ships] that the Martian people may have constructed. (*Argus*, 10 September 1892).

> Professor Pickering, of Peru, a charming embodiment of alliterative astronomical perspicacity, has discovered two ranges of mountains and eleven large lakes in Mars, and the director of the observatory of Nice [Henri Joseph Anastase Perrotin; 1845–1904], who has also taken a hand in the game, "comes in" with an assertion that he has seen "certain brilliant points on the edges of the planet." Professor Pickering, who, as an American and a lecturer at Harvard, is not likely to be easily bluffed, may be expected to "see" the director, and "go one better." Two ranges of mountains and eleven large lakes make up a very fair hand, but the professor's many friends will be disappointed if he does not pick up something still better at the next attempt. Three town-halls, four hopper barges, and a tramway-shed would make the game practically a certainty for him, although, of course, one never can tell what a Frenchman is equal to in an emergency. (*Gippsland Daily News*, 12 September 1892).

Nevertheless, the editor wistfully concludes, “However, the contest, inasmuch as it partakes of an international character, is bound to be exciting, and one cannot help wishing that Mr. Ellery would consent to take a seat at the same sidereal table for the honor [*sic*] of Australia.” In multiple newspaper reprints, Ellery suggested that brief cable messages were prone to exaggeration by correspondents, a point that aligned closely with his broader scepticism towards the press-driven amplification of ambiguous observations:

> Putting all these things together, and the statement in the cable message, and knowing that Professor Pickering had no instruments there such as are in the Northern Hemisphere, Mr. Ellery thought that when the facts became known it would be found that some bare statement of Professor Pickering's had been enlarged upon. He did not think that the markings on Mars were ever sufficiently clear under any circumstances to induce a man like Professor Pickering to make such a definite statement as that contained in the cable message. (*Argus*, 5 September 1892).

Ellery also played a prominent role in debates about the Martian atmosphere. When cable reports from the Lick Observatory suggested that Mars lacked an atmosphere altogether, Ellery countered with arguments grounded in seasonal polar-cap behaviour, asserting that the observed waxing and waning of polar whiteness implied the presence of water and, by extension, an atmosphere (e.g., *Argus*, 21 August 1894; *Brisbane Courier*, 27 August 1894; and subsequent reprints): “‘Where there is snow there must he water,’ says Mr. Ellery, ‘and where there is water there must be [an] atmosphere.’”[9] Although he acknowledged uncertainty in atmospheric density and composition, his insistence on empirical consistency positioned Australian astronomy as a corrective force within an increasingly polarised international debate.

Notably, Ellery’s interventions were frequently framed by newspapers as restoring calm to public discussion. Several reports explicitly credited him with relieving popular anxiety generated by exaggerated claims of Martian engineering or civilisation. In this sense, Ellery’s authority operated not merely at the level of scientific adjudication but also as a cultural stabiliser, reinforcing a vision of astronomy as disciplined inquiry rather than imaginative spectacle.

### 5.4 Numerical exaggeration and the rhetoric of proximity

Popular reporting on Mars frequently employed numerical claims intended to emphasise the planet’s unusual proximity during opposition, but these figures were not always astronomically accurate. In fact, one Australian headline confidently assured readers that Mars was “… AT CLOSE RANGE. EASY TO STUDY HIS FACE WHEN ONLY 35 MILLION MILES AWAY” (*Launceston Examiner*, 10 July 1899; *Maitland Daily Mercury*, 11 July 1899; *Australian Town and Country Journal*, 19 August 1899). Newspapers sometimes stated that Mars approached Earth to “within thirty million miles” or comparable distances, values that either underestimated the true minimum separation (~56 million km) or confused miles and kilometres. Such numerical exaggeration was not unique to the Australian press, but it was widely reproduced there through the reprinting of overseas material.

Even when some writers acknowledged the vastness of interplanetary distances, numerical framing remained misleading. One account of the 1877 opposition described Mars as “at his nearest to us … at the enormous distance of 35 million miles” (56 million km; *Launceston Examiner*, 10 July 1899), implicitly treating this value as an exceptional limit rather than a physical boundary. In some cases, numerical exaggeration appears to have arisen through the compression or careless transmission of figures. Several reports stated that Mars’s distance was “but 34 million miles” (54.7 million km; e.g., *W.A. Record*, 12 November 1891), including some that attributed the exaggerated distance to Russell himself, conflating minimum separation with broader orbital distances and thereby exaggerating the planet’s accessibility:

> One of its nearest oppositions will take place on the 5th of September next [1877], on which occasion Mars will then appear sixty times larger than when at its greatest distance from the [E]arth—the nearest approach being 34,000,000 miles [54.7 million kilometres] … (*Sydney Morning Herald*, 11 and 24 August 1877; *Sydney Mail and New South Wales Advertiser*, 18 August 1877; *Grenfell Record and Lachlan District Advertiser*, 8 September 1877)

Although such somewhat exaggerated figures were nevertheless largely grounded in genuinely favourable oppositions, their rhetorical use exaggerated what these distances made observationally possible, reinforcing the impression that Mars had been brought within unprecedented observational reach. This applied particularly during late-August and September oppositions, which were both closer and more favourably placed for southern observers. Whereas these favourable configurations did improve visibility, especially from the southern hemisphere, they did not resolve the interpretive uncertainties associated with telescopic observation. By highlighting dramatic proximity, newspapers conveyed the idea that Mars was temporarily rendered accessible to human scrutiny, encouraging readers to believe that the limits of telescopic observation were being pushed. In this context, numerical precision was secondary to narrative effect: distance functioned as a proxy for observational credibility. Australian editors rarely corrected such figures explicitly, but they often balanced them with qualifying language, noting uncertainty, disagreement among astronomers or observational difficulties. This pattern suggests that local newspapers acted as moderators rather than correctors: they preserved the rhetorical impact of proximity claims while coupling them with scepticism about their precision.

When Australian professional astronomers were cited directly, exaggerated numerical proximity was rarely challenged head-on, but it was frequently softened by explicit expressions of uncertainty. In Australian-written explanatory pieces, numerical proximity was often accompanied by explicit editorial hedging, thereby moderating proximity rhetoric without dismantling it.

### 5.5 Australian observational contributions: Walter F. Gale

In contrast to the institutional restraint of Ellery and Russell, Australian engagement with Mars was occasionally advanced through the work of skilled amateur observers (see, e.g., Figure 6). Among Australian observers, Walter Frederick Gale (1865–1945; see McCarthy, 1993–2018) of Paddington (Sydney), later manager of the Newcastle branch of the Savings Bank of New South Wales, occupies a distinctive position in Australian newspaper reporting. Gale, depicted in Figure 7, appeared in the press primarily as an observer and illustrator of Mars. His sketches of the planet's surface (see Figure 8 for examples) were reproduced or described in newspaper articles, lending visual immediacy to discussions that were otherwise dominated by textual summaries of overseas observations. We learn from the *Newcastle Morning Herald and Miners' Advocate* of 8 February 1897 that

> [i]n 1890 and '92 he happened to be in possession of a very fine reflecting telescope of 8¼ inches [21 cm] aperture, giving roughly a thousand times the light grasp of the unaided eye; and he says it was as easy sometimes to make out the oceans and continents of this distant world as to see the configuration of the [E]arth shown upon a globe a few yards off.

These illustrations reinforced the credibility of Mars reporting by demonstrating that meaningful observation was possible from Australia and by Australians, particularly under favourable southern-hemisphere viewing conditions.

Gale's observations were noted beyond the colonial context, and his suggestion that certain canals resolved into discontinuous or lake-like features was taken up in wider discussions of the canal question, anticipating later sceptical interpretations. As a case in point, these suggestions were cited by Edward Walter Maunder (1851–1928) of the Royal Observatory, Greenwich, as a possible indication of a new phase in the canal debate and anticipated later sceptical interpretations associated with observers such as Vincenzo Cerulli (1859–1927). An article by Maunder, published in the *London Daily Graphic* on 15 December 1892, was illustrated with Gale's drawings of Mars, explicitly attributing their clarity to the planet's advantageous placement for southern observers (cited in the *Sydney Morning Herald*, 2 February 1893). This episode represents one of the clearest instances in which Australian-produced Martian imagery and occasional discoveries of new surface features (*Newcastle Morning Herald and Miners' Advocate*, 8 February 1897) entered metropolitan popular astronomy, mediated through an authoritative British institution:

> In the same month [August 1894] four new lakes lying very near the [Martian] equator, to the east of *Trivium Charontis* ['Crossroad of Charon', a dark surface feature in Mars's Elysium region], were discovered at Paddington. The lakes were seen and confirmed at the great Lowell Observatory in Arizona during 1894, and it has been ascertained that the eminent French astronomer now confirms an important observation of the same night. (*Ibid.*)

Gale's subsequent participation in the Lick Observatory solar-eclipse expedition to Peru in 1893 (*Sydney Morning Herald*, 2 February 1893) further reinforced his standing as an observer capable of contributing to internationally coordinated scientific enterprises.

Importantly, Gale's contributions remained observational rather than interpretive. Whereas his drawings depicted linear features and surface detail, they were not accompanied by claims regarding artificiality or intelligent design. In public lectures and press statements he explicitly rejected the notion that the so-called canals were engineered works, invoking their immense aggregate length and breadth to argue against artificial origin:

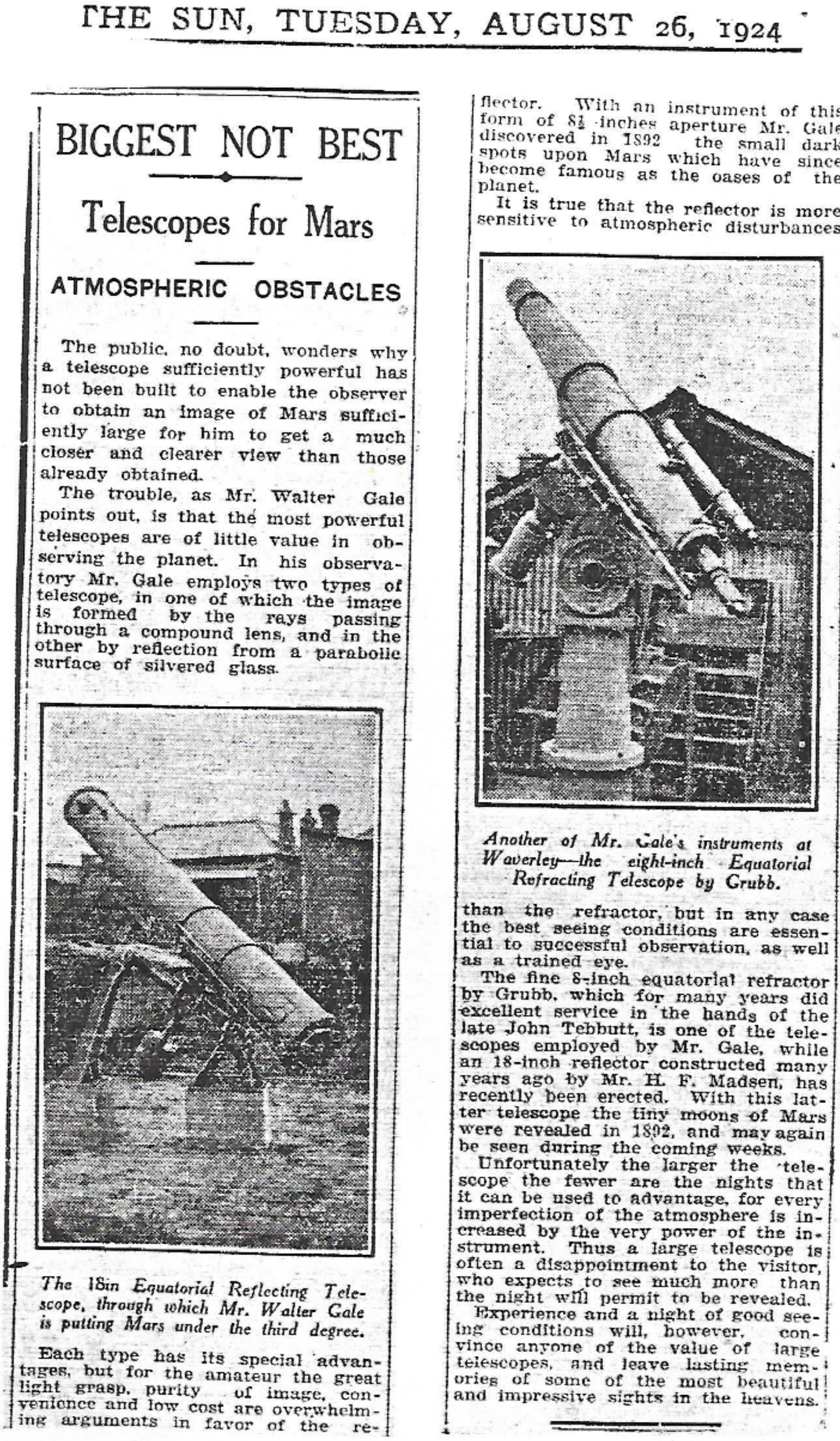

THE SUN, TUESDAY, AUGUST 26, 1924

# BIGGEST NOT BEST

## Telescopes for Mars

### ATMOSPHERIC OBSTACLES

The public, no doubt, wonders why a telescope sufficiently powerful has not been built to enable the observer to obtain an image of Mars sufficiently large for him to get a much closer and clearer view than those already obtained.

The trouble, as Mr. Walter Gale points out, is that the most powerful telescopes are of little value in observing the planet. In his observatory Mr. Gale employs two types of telescope, in one of which the image is formed by the rays passing through a compound lens, and in the other by reflection from a parabolic surface of silvered glass.

*The 18in Equatorial Reflecting Telescope, through which Mr. Walter Gale is putting Mars under the third degree.*

Each type has its special advantages, but for the amateur the great light grasp, purity of image, convenience and low cost are overwhelming arguments in favor of the reflector. With an instrument of this form of 8½ inches aperture Mr. Gale discovered in 1892 the small dark spots upon Mars which have since become famous as the oases of the planet.

It is true that the reflector is more sensitive to atmospheric disturbances than the refractor, but in any case the best seeing conditions are essential to successful observation, as well as a trained eye.

*Another of Mr. Gale's instruments at Waverley—the eight-inch Equatorial Refracting Telescope by Grubb.*

The fine 8-inch equatorial refractor by Grubb, which for many years did excellent service in the hands of the late John Tebbutt, is one of the telescopes employed by Mr. Gale, while an 18-inch reflector constructed many years ago by Mr. H. F. Madsen, has recently been erected. With this latter telescope the tiny moons of Mars were revealed in 1892, and may again be seen during the coming weeks.

Unfortunately the larger the telescope the fewer are the nights that it can be used to advantage, for every imperfection of the atmosphere is increased by the very power of the instrument. Thus a large telescope is often a disappointment to the visitor, who expects to see much more than the night will permit to be revealed.

Experience and a night of good seeing conditions will, however, convince anyone of the value of large telescopes, and leave lasting memories of some of the most beautiful and impressive sights in the heavens.

**Figure 6.** Walter F. Gale's rise to prominence as leading amateur astronomer in the colonies. (*The Sun*, 26 August 1924: p. 14; no known copyright restrictions.)

In Mr. Gale's opinion the canals of Mars are not canals in the sense we employ the word at all. … The idea of artificial origin must, Mr. Gale thinks, be put on one side, unless the Martians greatly excel the inhabitants of the [E]arth in engineering. (*Newcastle Morning Herald and Miners' Advocate*, 8 February 1897).

In this respect, Gale's work aligns more closely with the restrained observational authority exemplified by Hall than with the interpretive ambition associated with Lowell (see Section 5.6).

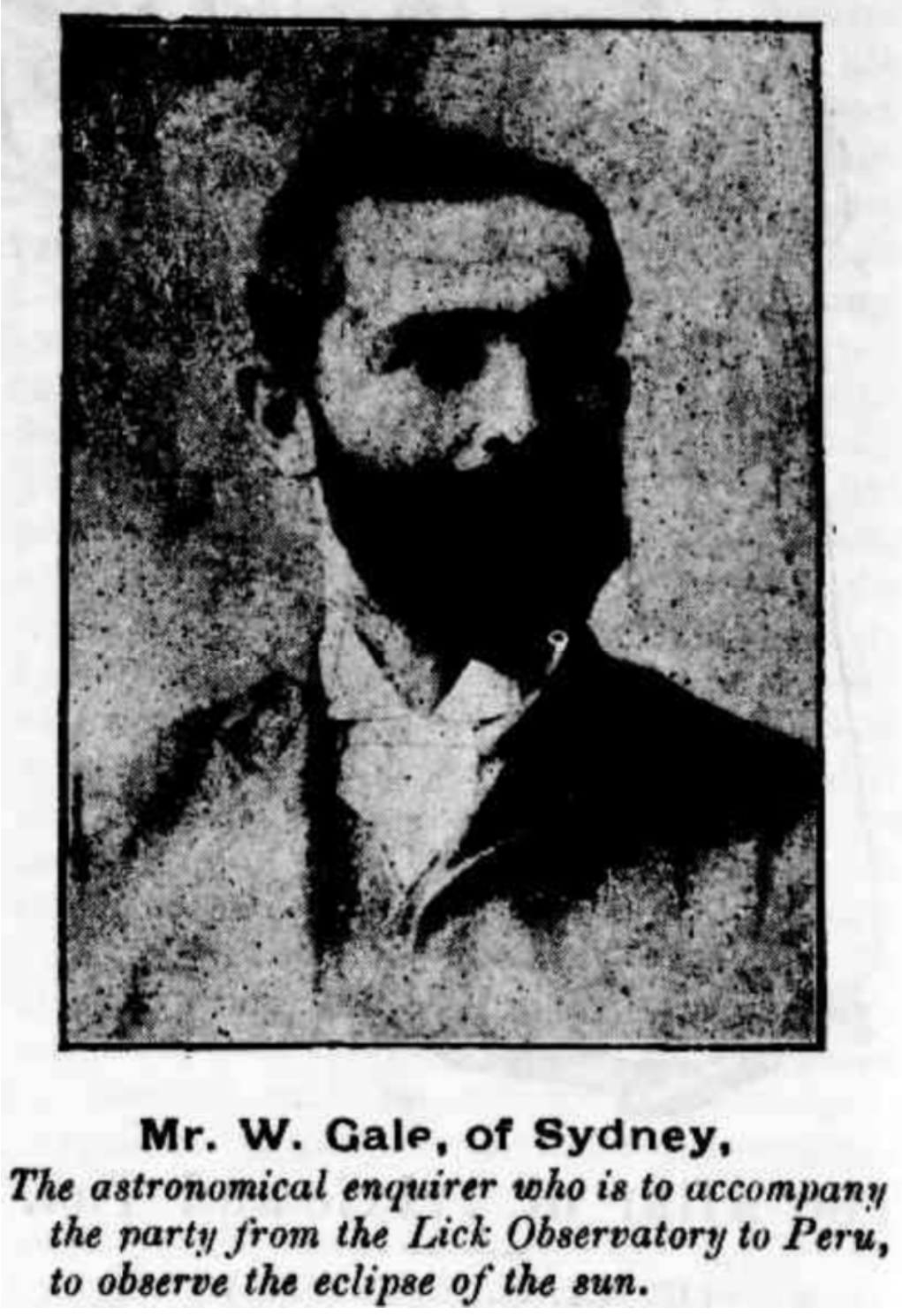

**Figure 7.** Walter F. Gale. (*Illustrated Sydney News*, 18 February 1893: p. 18; no known copyright restrictions.)

The press treatment of Gale thus highlights a distinct mode of Australian participation in Mars astronomy. Rather than shaping theoretical debate, Australian observers contributed through confirmation, illustration and local engagement. Their work validated overseas claims without driving speculative escalation. Gale's presence in the Australian Mars narrative therefore reinforces a key theme of this paper: that while Mars excitement was widely shared, interpretive leadership remained largely external, with Australian contributors occupying epistemically conservative yet observationally authoritative roles within an international framework of observation (e.g., *Newcastle Morning Herald and Miners' Advocate*, 8 February 1897). That Gale's drawings circulated beyond Australia, appearing in British publications and informing overseas discussions, underscores the permeability of colonial and metropolitan scientific boundaries.

### 5.6 American prominence: Pickering and Lowell

Pickering occupies a transitional position in Australian newspaper coverage. His association with the Arequipa observatory high in the Peruvian mountains, established to take advantage of favourable southern-hemisphere viewing conditions, aligned him with systematic and purposeful observation (e.g., Pickering, 1892; Grassie, 1897). Reports frequently highlighted the geographical rationale for the observatory, reinforcing the idea that Mars demanded specialised observational strategies:

> The transparency of the air [at Arequipa] is most remarkable, but the steadiness is such as to be almost beyond belief. In fact, this, the chief difficulty of the modern astronomer, is here practically entirely overcome. This we attribute in part to our altitude, but very largely to the excessive dryness of our climate. … It happens, by a curious coincidence, that many of the most interesting bodies in the whole sky cluster around the southern pole of the heavens, and are, therefore, never visible in the United States or Europe. Those objects are all well seen from Arequipa. In fact, an area equal to one quarter of the whole sky can never be studied to advantage at any northern observatory. (*Ibid.*)

Pickering's observations of Martian surface features and atmospheric phenomena were often reported with cautious interest. Although he did not initially promote strong claims regarding artificiality, his work expanded the observational foundation upon which such interpretations could be built. In press narratives, Pickering thus appears as a figure who broadened the scope of Martian observation while maintaining a posture of scientific seriousness. This positioning made his work particularly amenable to later reinterpretation and reprinting. As a case in point, we read in the *South Australian Register* of 1 October 1892:

> Professor Pickering, we have been informed in the papers, has been making some wonderful discoveries, having seen two mountain ranges and eleven lakes. We must, however, place this report in the same category as Schiaparelli's canals, and suspend our judgment whilst awaiting definite confirmation. I may say that Arequipa, where Professor Pickering is situated, possesses the best known atmospheric conditions for observing in the world.

The most dramatic shift in Mars reporting occurred with the increasing prominence of Lowell in the mid-to-late 1890s. Lowell's influence rested not only on observational claims but on an unusually effective media strategy that combined observatory reports, public lectures, illustrated journalism and transatlantic syndication. Australian newspapers presented Lowell not merely as an observer but as an interpreter of Mars, offering coherent and confident explanations for the planet's surface features.

Lowell’s authority derived not only from his observational claims but from his willingness to synthesise those observations into a comprehensive theory, e.g.:

> Mr. Lowell, of the Lowell Observatory, who has been making continued and careful observations of the planet Mars, says of the mysterious canals:—"Were they more generally observable, the world would have been spared much scepticism and more theory. They may, of course, not be artificial, but observations here indicate that they are, as will, I think, appear from the drawings. For it is one thing to see two or three canals, and quite another to have the planet’s surface mapped with them upon a most elaborate system of triangulation.” (*Morning Bulletin*, 31 December 1894).

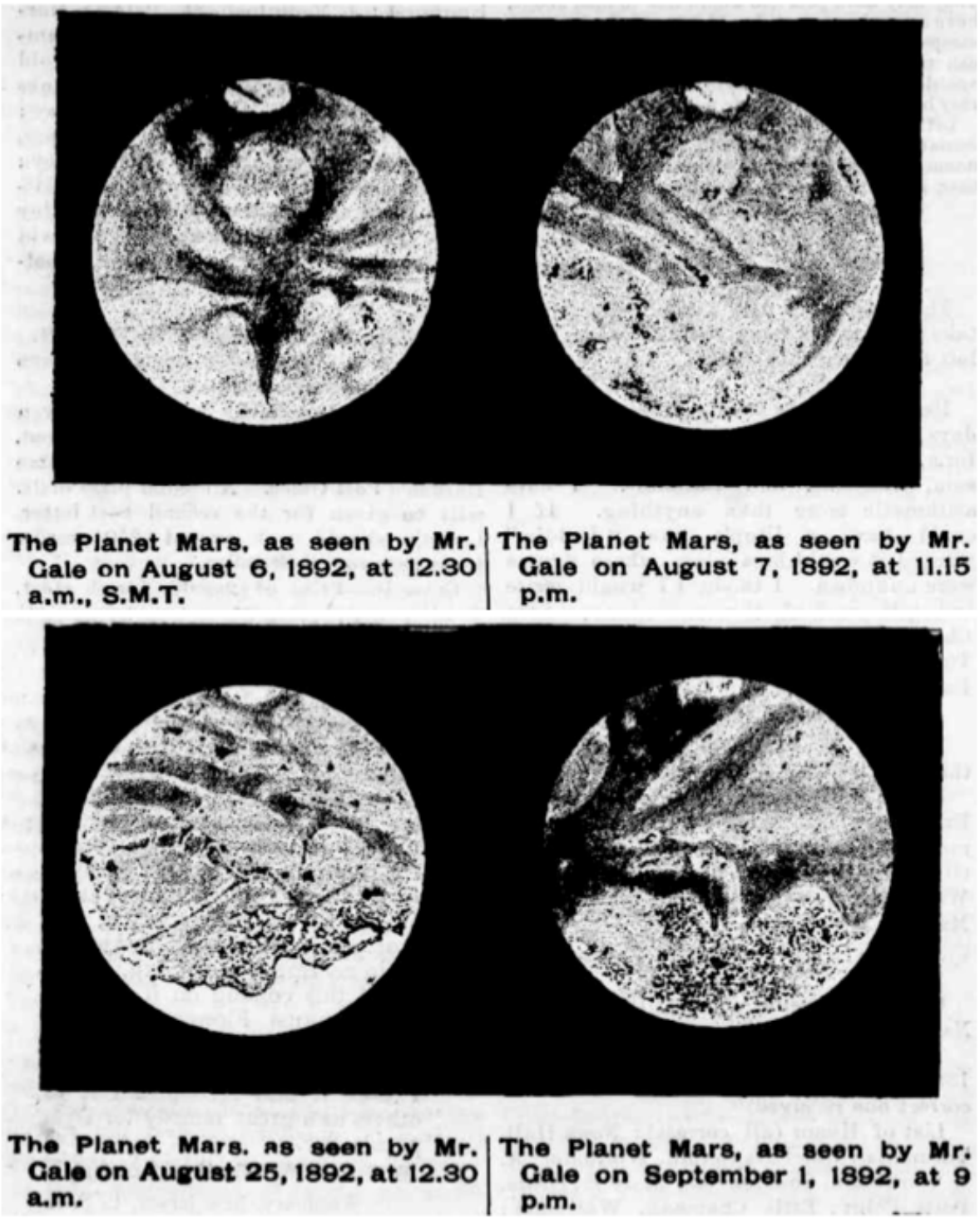

**Figure 8.** Original drawings of Mars by Walter F. Gale. (*Illustrated Sydney News*, 18 February 1893: p. 18; no known copyright restrictions.)

His arguments concerning the artificial nature of Martian canals, planetary desiccation and intelligent adaptation were often summarised in extended articles that conveyed a sense of narrative completeness (e.g., *The Australasian*, 12 October 1895; *Argus*, 16 November 1895; *Maitland Daily Mercury*, 4 February 1896; and numerous reprints). In contrast to earlier reporting, which balanced implication with uncertainty, Lowell-associated articles frequently emphasised interpretation. While sceptical voices were sometimes acknowledged, the clarity and assertiveness of Lowell’s views gave them particular traction in the press. An example of overseas drawings of Martian features reproduced in Australian newspapers is shown in Figure 9.

### 5.8 Context

Australian newspapers generally maintained a degree of editorial distance from Lowell’s most ambitious claims. Rather than endorsing his conclusions outright, they attributed them carefully, allowing readers to engage with the ideas while preserving journalistic neutrality, e.g.:

> Against all which, to mention nothing else, stands the fundamental doubt whether so small a globe as Mars, with so rare an atmosphere, and receiving from the [S]un only half as much heat to each square mile as does the [E]arth, can possibly maintain anywhere a temperature even as high as that which prevails on the summits of our loftiest mountains; whether, in fact, the polar-caps are made of frozen water or of some very different substance. (*Maitland Daily Mercury*, 4 February 1896).

The editorial handling of Mars-related material in Australian newspapers further reveals an active, rather than neutral, mode of engagement. Reports frequently covered competing interpretations, emphasised uncertainty or explicitly noted disagreement among astronomers. This sceptical filtering did not diminish public interest; instead, it framed Mars as an unresolved scientific problem. By moderating speculative enthusiasm while preserving curiosity, Australian editors acted as gatekeepers of scientific plausibility, shaping the tone and limits of public engagement with Martian ideas. This strategy enabled newspapers to capitalise on the appeal of speculative interpretation without undermining their credibility. Lowell’s prominence thus illustrates how individual authority could amplify interpretive excitement while remaining compatible with journalistic norms of caution.

The increasing centrality of individual figures also introduced a more overtly contested dimension to Mars reporting. Disagreements between astronomers, differing interpretations of observational evidence and debates about optical illusion versus physical reality became newsworthy in their own right, e.g.:

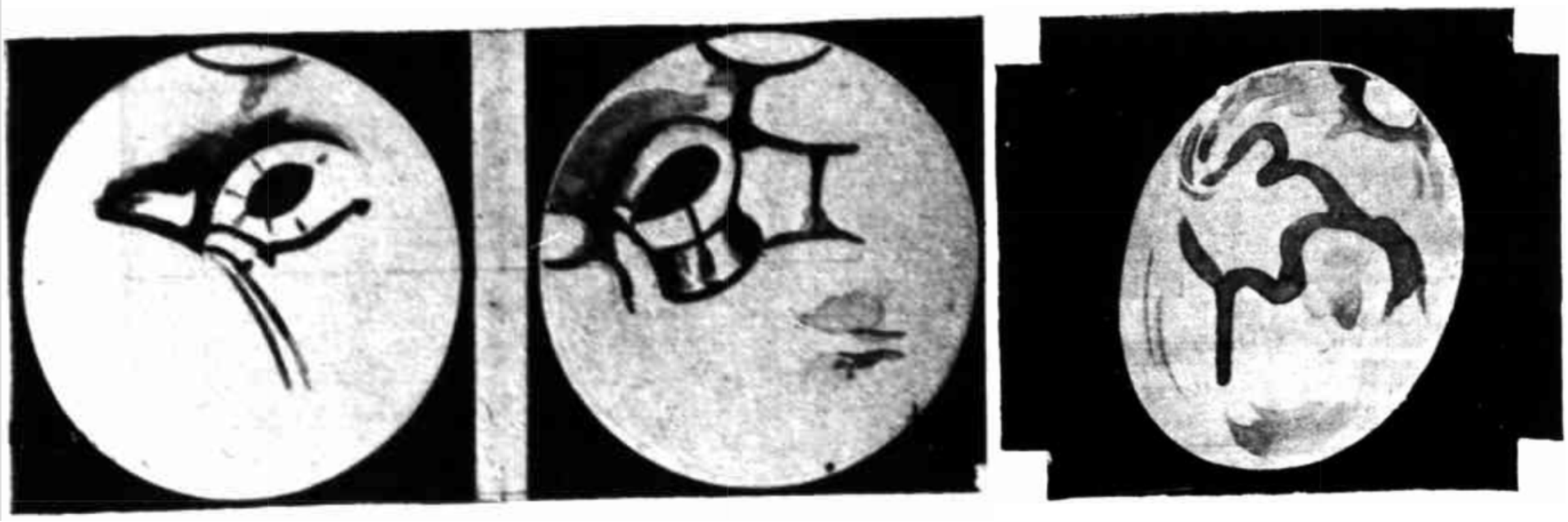

**Figure 9.** Examples of drawings of Martian surface features reproduced in Australian newspapers. (*The Australasian*, 6 April 1895: p. 28; no known copyright restrictions.)

> But many investigators, notably Professor Lowell, go farther. The Professor points to the extreme regularity of the canals; ... This implies, in his judgment, a corresponding regularity in the cutting of the watercourses, a regularity which shows them to be works of art. (*Riverine Herald*, 26 October 1896).

Such controversies further sustained public interest, transforming Mars from a distant object of study into an arena of scientific debate. Similar disagreements unfolded contemporaneously in Britain and the USA, where figures such as Lockyer, Pickering and Lowell publicly contested both observational reliability and interpretive legitimacy. The press capitalised on these disagreements by simultaneously covering contrasting viewpoints, reinforcing the sense that Mars was a subject of active and unresolved inquiry.

The evolution of Mars reporting in the Australian press reflects a broader shift from institutional to personality-driven authority. Whereas early coverage emphasised collective observation and restraint, later reporting increasingly revolved around named individuals whose interpretations shaped public understanding. Hall, Pickering and Lowell represent distinct modes of authority—observational, organisational and interpretive—each contributing differently to the Australian press narrative. Their contrasting roles illustrate how scientific personalities could guide, expand and intensify the meanings attached to Mars, setting the stage for the more speculative and imaginative engagements examined in the following section.

The cautious public posture adopted by Russell and Ellery stands in marked contrast to the media strategy later pursued by Lowell. Whereas Australian government astronomers sought to control the timing and framing of disclosure, often resisting press demands for immediate interpretation, Lowell actively cultivated publicity as an integral component of his scientific programme. Through lectures, interviews and vividly written popular works, Lowell treated public engagement not as a secondary obligation but as a means of shaping scientific legitimacy itself. This marks a structural shift in the relationship between astronomy and the press: from an older model in which authority derived from institutional restraint and delayed publication, to a newer mode in which visibility, repetition and narrative coherence became central to the reception of astronomical claims. Australian newspaper responses to Lowell's assertions, when read against earlier frustration with Russell's reticence (Section 5.2), reveal a press increasingly willing to reward interpretive confidence over observational reserve, even as doubts about evidentiary sufficiency persisted.

## 6. LIFE, INTELLIGENCE AND PLANETARY FUTURES

### 6.1 From surface features to living worlds

As interpretations of Martian surface features gained confidence in the Australian press, attention increasingly turned from physical description to biological and social implication. The question was no longer merely what existed on Mars, but what kind of world it represented. Reports began to explore the possibility that the planet supported life, often framing the issue as a natural extension of observational findings rather than a speculative leap. This transition was facilitated by the cumulative authority established in earlier reporting. Once canals were accepted as plausible features and engineering analogies had been normalised, the inference of intelligent agency became easier to articulate.

A further contribution of Australian Mars reporting lay in its emphasis on visual mediation. Newspapers frequently reproduced sketches, described illustrations or provided detailed verbal guidance intended to help readers imagine or recognise Martian features. As a case in point, the *London Daily Graphic* of 15 December 1892 is said to be "… illustrated with sketches made by Mr. W. F. Gale at his observatory at Paddington, which show the canals, &c., to great advantage, on account of Mars having been well placed for southern observers" (*Sydney Morning Herald*, 2 February 1893). Such material encouraged a participatory mode of engagement, inviting readers to 'see' Mars through the eyes of observers rather than merely accept abstract descriptions. This visual framing reinforced the observational character of Mars astronomy and helped bridge the gap between distant telescopic practice and everyday readership. In doing so, Australian newspapers contributed not only to what readers knew about Mars but to how they were encouraged to visualise it.

A recurring theme in late-nineteenth-century Mars reporting was the idea that Mars represented an advanced stage of planetary evolution. Martian intelligence was often portrayed as adaptive. Canals were interpreted not as symbols of expansion or conquest, but as evidence of a civilisation managing decline. This narrative carried a moral dimension, suggesting foresight, cooperation and technological ingenuity in the face of environmental constraint. Discussions of Martian life in the Australian press tended to emphasise scale and coordination rather than individual agency. Martians were imagined as collective actors, capable of undertaking planet-wide projects that exceeded the capacities of human societies. This emphasis reinforced the plausibility of canals as infrastructure while distancing Martian intelligence from human-like behaviour. Such portrayals avoided anthropomorphism in the narrow sense, even as they attributed rationality and purpose to an alien civilisation. Mars thus became a mirror in which human technological aspirations and anxieties could be reflected without invoking direct analogy.

### 6.2 Scientific caution and imaginative openness

Despite the growing confidence of interpretive narratives, Australian newspapers continued to assert the provisional nature of claims about Martian life. Articles often included explicit acknowledgements of uncertainty, invoking alternative explanations or reminding readers of observational limitations. This balance between imaginative openness and scientific caution was a defining feature of Mars reporting. It allowed newspapers to engage readers' curiosity while maintaining alignment with scientific norms. Speculation was framed as reasonable and informed, not sensational or fantastical.

In some instances, discussions of Martian intelligence took on broader ethical or philosophical significance. Mars was invoked as a thought experiment through which questions of progress, sustainability and planetary stewardship could be explored. The image of a civilisation confronting environmental limits resonated with contemporary concerns about resource use and technological responsibility, particularly in Australia. Although such reflections were typically implicit rather than explicit, they contributed to the depth and persistence of Mars-related discourse in the press. Notably, Australian newspapers generally refrained from portraying Martians as hostile or invasive. The emphasis remained on observation, adaptation and engineering rather than conflict. This restraint distinguishes press reporting of Mars in the nineteenth century from later fictional representations and underscores the scientific framing that continued to dominate. Mars was imagined as inhabited, perhaps by intelligent residents, but fundamentally distant and observationally bounded.

By the late 1890s, Australian newspaper reporting had transformed Mars from a distant planet into a speculative living world. Discussions of life and intelligence emerged organically from earlier observational and interpretive frameworks, supported by analogies of engineering, adaptation and planetary ageing. Yet these narratives remained constrained by scientific caution, maintaining clear boundaries between evidence and inference. For instance, Ellery's repeated rejection of canal-as-irrigation models offers a good example of Australian institutional resistance to speculative convergence. This balance set the stage for the final phase of my analysis, which examines how these press-based interpretations related to contemporary fiction and popular culture and where the limits of journalistic imagination lay.

## 7. POPULAR CULTURE, FICTION AND THE LIMITS OF INFLUENCE

### 7.1 Speculative Mars before fiction

By the final decade of the nineteenth century, Australian newspaper readers were already familiar with many of the themes that would later dominate fictional treatments of Mars. Canals, planetary desiccation and intelligent adaptation had become established elements of press discourse, grounded in scientific reporting and attributed to recognised authorities. This pre-existing speculative framework meant that Mars did not require fictional invention to become culturally resonant. Instead, newspapers themselves functioned as vehicles for imaginative engagement, operating within the bounds of scientific plausibility.

Despite the increasing richness of interpretive narratives, Australian newspapers generally maintained a clear distinction between scientific reporting and imaginative speculation. Even when discussing intelligent life on Mars, articles were careful to anchor claims in observational evidence and authoritative attribution. This restraint shaped the tone of Mars reporting. The planet was portrayed as intriguing and suggestive but not sensational. Speculative possibilities were framed as topics for reasoned consideration rather than dramatic narrative, reinforcing the press's role as mediator rather than storyteller.

### 7.2 H. G. Wells and *The War of the Worlds*

Wells's *The War of the Worlds*, serialised in 1897 and published in book form in 1898, occupies a prominent place in later accounts of Martian imagery. However, its direct impact on Australian newspaper reporting during our period of interest appears limited. References to Wells's work in Australian newspapers prior to 1899 are relatively sparse and typically confined to literary notices rather than scientific commentary. Where mentioned, the novel is treated explicitly as fiction, distinct from observational or interpretive discussions of Mars. This separation underscores the extent to which press-based Mars excitement developed independently of fictional influence.

The Australian press cultivated what might be termed a *scientific imagination*: a mode of engagement that allowed readers to contemplate extraordinary possibilities while remaining anchored to empirical discourse. Mars served as a particularly effective site for this mode of thought, combining observational accessibility with interpretive openness. Unlike fiction, which could resolve uncertainty through narrative, newspaper reporting sustained ambiguity. This open-endedness kept Mars provisional, inviting repeated engagement without closure. The contrast between journalistic and fictional treatments of Mars highlights the boundaries maintained within late nineteenth-century public discourse. Although both drew on similar imagery and themes, their purposes and conventions differed. Newspapers prioritised credibility, attribution and restraint; fiction prioritised narrative coherence and dramatic effect. Australian press coverage of Mars thus cannot be understood simply as a precursor to later fictional imaginings. Instead, it represents a distinct and historically specific mode of engagement with planetary science.

The relationship between Mars reporting and popular culture in late nineteenth-century Australia was one of proximity without convergence. Scientific journalism generated rich interpretive possibilities, but it stopped short of the dramatic narratives that would later dominate popular imagination. By maintaining this distinction, newspapers preserved Mars as a legitimate subject of scientific inquiry while allowing readers to explore its implications. This balance marks the culmination of the interpretive trajectory traced in this paper.

## 8. INTERPRETING MARS IN COLONIAL AUSTRALIA

### 8.1 Mars as a mediated scientific object

The analysis presented in this paper demonstrates that Mars, as encountered by Australian newspaper readers between 1875 and 1899, was a fundamentally mediated object. Observations made at distant observatories were transformed through journalistic language, analogy and attribution into narratives that were intelligible, engaging and open to interpretation. This mediation did not distort scientific knowledge so much as reframe it. Newspapers selected and emphasised specific aspects of Mars-related astronomy—surface features, canals, habitability—while subordinating others. In doing so, they shaped the parameters within which readers could imagine the planet and its possibilities.

Australian newspapers also played a role in sustaining interest in Mars beyond moments of heightened observational opportunity. Through reprinting, delayed circulation and regional dissemination, Mars-related material continued to appear well after initial publication. This temporal persistence transformed Mars from an episodic curiosity into an enduring subject of public attention. In maintaining the planet's visibility between oppositions, the Australian press contributed to the stabilisation of Mars as a continuing scientific and imaginative concern.

One of the most striking features of Australian Mars reporting is the degree of interpretive richness achieved without recourse to sensationalism. Even when discussing intelligent life or planet-wide engineering works, newspapers generally adhered to conventions of scientific restraint. Claims were attributed, uncertainty was acknowledged and speculative conclusions were framed as provisional. This approach contrasts with later popular representations of Mars and highlights the distinctiveness of late nineteenth-century science journalism. The press enabled imaginative engagement while preserving epistemic caution, producing a discourse that was speculative yet credible.

Authority played a central role in determining which interpretations gained traction. Named figures such as Hall, Pickering and Lowell provided focal points around which meaning could coalesce, each embodying a different relationship between observation and interpretation. Lowell's prominence, in particular, illustrates how interpretive confidence could reshape press narratives. His willingness to synthesise observational data into a coherent theory allowed newspapers to move beyond cautious description towards more expansive speculation, while still grounding claims in recognised expertise.

Although Australian Mars reporting relied heavily on overseas sources, it unfolded within a distinctive colonial context. Environmental conditions such as aridity, irrigation challenges and agricultural expansion provided an unspoken backdrop against which discussions of Martian canals and planetary decline resonate. Mars could thus be read not only as a distant world but as a site for reflecting on environmental management and technological adaptation. This resonance does not imply direct analogy, but it helps explain why certain interpretations proved particularly compelling.

The relative absence of interpretive commentary from Australian institutional astronomers such as Ellery and Russell further underscores the press-driven nature of Mars enthusiasm. Whereas these figures lent authority to astronomy more broadly, they did not act as catalysts for speculative engagement with Mars, leaving newspapers to negotiate the planet's meanings largely through overseas voices. The observational contributions of figures such as Gale further underscore the asymmetry between Australian participation and interpretive authority: Australians observed Mars but rarely claimed to explain it.

### 8.2 Mars before modern science fiction

The results of this study complicate narratives that position science fiction as the primary driver of Martian imagery. In Australia, at least, many of the themes later popularised by fiction were already firmly embedded in scientific journalism before the end of the nineteenth century. This does not diminish the importance of literary works such as *The War of the Worlds*, but it situates them within an existing interpretive landscape shaped by press reporting. Mars was already a world of canals, decline and intelligence before it became a world of invasion and catastrophe.

Read alongside Paper 2, which examines the circulation and temporal structure of Mars reporting, this paper highlights the interpretive consequences of journalistic mediation. The mechanisms that enabled Mars-related material to circulate widely also shaped how it was understood. Together, the two papers demonstrate that meaning and mechanism are inseparable in the public history of astronomy. The Australian case illustrates the importance of considering colonial press environments in the history of science. Far from being passive recipients of metropolitan knowledge, colonial newspapers actively participated in the construction of scientific meaning, e.g.:

> As all the world knows, the planet Mars has for years been mapped out into seas, continents, and islands by the astronomers, who appear to be as much at home on the surface of the red planet as in [Brisbane's] Queen-street … (Endymion, 1882).

Mars, with its combination of observational uncertainty and interpretive openness, provides an especially revealing example. Through the press, planetary science became part of everyday intellectual life, engaging readers in questions that extended beyond astronomy to encompass progress (e.g., Stella, 1881; Sirius, 1882a,b; Gerrard, 1892), environment and the future of civilisation. Australian contributions to Mars excitement thus operated less through theoretical innovation than through observation, visualisation, editorial judgement and temporal persistence.

## 9. CONCLUDING THOUGHTS

Between 1875 and 1899, Mars occupied a distinctive and evolving place in the Australian press. Initially presented as a distant object of careful observation, the planet gradually became a site of interpretive engagement, shaped by analogies of engineering, environmental adaptation and technological intelligence. This transformation occurred through the cumulative effects of repeated reporting, authoritative attribution and cautious speculation. Australian newspapers played an active role in this process. By selecting, framing and reprinting overseas scientific material, they shaped how Mars was understood by colonial readers. Observations were not merely transmitted; they were contextualised, debated and given meaning. The press thus functioned as a crucial intermediary between astronomical practice and public imagination. Russell and Ellery, in turn, played their roles as active agents in shaping Australian astronomical discourse, not merely as conduits for overseas ideas.

The interpretation of Martian canals provides a particularly revealing lens through which to view this process. What began as ambiguous observational features acquired interpretive significance through language, analogy and repetition. Engineering metaphors, planetary ageing narratives and appeals to scientific authority allowed newspapers to explore the possibility of intelligent life while maintaining epistemic restraint. Importantly, this interpretive richness developed largely independently of fictional influence. By the end of the nineteenth century, many of the themes later associated with science fiction were already present in scientific journalism (e.g., *Albury Banner and Wodonga Express*, 12 October 1883). Mars was imagined as inhabited, adaptive and technologically capable long before it was imagined as hostile or invasive.

This study contributes to the history of astronomy by highlighting the interpretive dimensions of press-mediated science in a colonial context. It demonstrates that Australian engagements with Mars were neither peripheral nor derivative but formed part of a global discourse shaped by local editorial practices and cultural resonances. Together with Paper 2, this study underscores the importance of examining both the mechanisms and meanings of scientific communication. Mars in the Australian press was not simply a reflection of astronomical discovery; it was a product of how science was written, circulated and read. Recovering this history enriches our understanding of the public life of astronomy at the close of the nineteenth century. By examining Australian Mars reporting within its international information ecosystem, this study demonstrates how colonial press cultures functioned as stabilising, filtering and legitimising agents in global scientific controversies rather than as passive recipients of metropolitan knowledge. In doing so, it invites further comparative studies of colonial scientific publics and their role in shaping the global reception of astronomical knowledge.

## 10. NOTES

1. For representative accounts, see Crowe (1986), Sheehan (1996), Crossley (2011) and Lane (2011). These monographs provide detailed analyses of Mars observation and interpretation but devote little attention to colonial or southern-hemisphere press cultures.

2. For discussions of the institutionalisation of Martian speculation around the turn of the twentieth century, particularly in relation to Lowell's observatory and publications, see Sheehan (1996) and Lane (2011). On the transformation of scientific communication through mass-market illustration and popular media after 1900, see Secord (2004), Lightman (2007) and Bowler (2009). The growing cultural influence of science fiction, especially following publication of Wells's *The War of the Worlds*, is analysed in Crossley (2011) and Stableford (1985).

3. Much of the late nineteenth-century excitement surrounding Mars can be traced to the mistranslation of Schiaparelli's term *canali*, introduced following his observations during the 1877 opposition. In Italian, *canali* denotes channels or grooves and carries no implication of artificial construction. Schiaparelli himself carefully avoided committing to any theory regarding the nature or origin of the linear features he described. However, in English-language reporting the term was routinely rendered as "canals", a word strongly associated with large-scale engineered waterways. Contemporary commentators, including J. Norman Lockyer (1836–1920) and Australian journalists, later acknowledged that this literal translation encouraged unwarranted assumptions about intelligent design and advanced civilisation on Mars. As several Australian newspapers observed in the early 1890s, the semantic slippage from natural channels to artificial canals played a major role in inflaming public speculation and giving an interpretive weight to Schiaparelli's observations that he had not intended. *The Age* of 21 January 1893 explained, "What we call a canal would not be visible at all on any of the planets, and the canals of Mars should never again be mentioned where English is spoken. The full absurdity of the mistake would be more obvious if we were to call the silver streak which separates France from England 'The English Canal'."

4. References to Mars's surface as "Holland"—in terms of the country's lack of significant geographic elevation—are found regularly in the Australian compilation, e.g., "… Mars… has reached the stage to which we are gradually tending, and has become one vast level plain. Wind and water have completed their work of denudation, wearing down the mountains and filling up the valleys, so that the whole surface resembles one vast Holland. At this point the raison d'etre of the canals comes in. The coast line of the land having been dammed against the sea, waterways have become an obvious convenience, if not a necessity of locomotion" (*Maitland Daily Mercury*, 18 October 1898; and subsequent syndication).

5. The Mildura irrigation experiment was a highly publicised late nineteenth-century attempt to transform semi-arid land along the Murray River in Victoria through large-scale canal engineering (for Martian parallels, see also the *Upper Murray and Mitta Herald*, 28 June 1888).

6. The “M. Henry” of Paris Observatory mentioned in Australian press reports refers to Paul Henry (1848–1905). Contemporary newspapers frequently did not distinguish between Paul and his brother, Prosper Henry (1849–1903), both of whom were senior astronomers at Paris Observatory and who were often cited simply as “MM. Henry” in telegraphic summaries. Prosper Henry was more strongly associated with instrumental development and astrophotography, whereas Paul Henry more frequently appears in connection with visual confirmations. Note that *Week* (10 November 1877) credits both Henry brothers for the confirmation: “… Messrs. Henry Brothers having observed the more distant of the satellites at the observatory of the latter town on the night of the 27th ult[imo].”
7. William Huggins’s pioneering spectroscopic study of Mars marked the first systematic attempt to investigate the planet’s atmosphere using spectral analysis. Examining the planet’s reflected solar spectrum, Huggins reported absorption features that he interpreted as evidence for an atmosphere containing water vapour. He argued that Mars’s reddish hue could not be explained solely by surface reflection. Although later observers questioned the sensitivity of early spectroscopes and the attribution of specific absorption lines to Martian atmospheric constituents, Huggins’s work exerted considerable influence on late nineteenth-century discussions of Martian habitability. His conclusions were frequently cited in popular and scientific discourse, including in Australian newspapers, as authoritative evidence that Mars possessed an atmosphere broadly comparable to—if thinner than—that of the Earth.
8. Cassiopeia is a prominent circumpolar constellation in the northern skies, composed of relatively bright, fixed stars whose positions change only on astronomical timescales. In the nineteenth century it featured routinely in positional astronomy, stellar cataloguing and calibration work; it was not associated with transient phenomena or speculative interpretation. Its invocation in press commentary therefore functioned as a rhetorical contrast with Mars, whose periodic oppositions, surface markings and alleged variability rendered it an object of exceptional public interest. The comparison implicitly criticised the application of observational reserve appropriate to routine stellar work to a planetary phenomenon that had, by the early 1890s, become embedded in popular scientific discourse.
9. Modern observations from orbiting spacecraft have confirmed that Mars’s white polar caps are composed of real ice deposits whose seasonal advance and retreat reflect atmospheric and climatic processes. The seasonal caps consist primarily of frozen $CO_2$ (carbon dioxide) that condenses from and returns to the atmosphere each year, while the residual (permanent) caps are dominated by water ice with varying thin layers of $CO_2$ ice depending on hemisphere and season. In the northern hemisphere most of the $CO_2$ frost disappears each summer, exposing nearly pure water-ice deposits, and in the south a thicker layer of $CO_2$ ice overlies an underlying water-ice cap (Kieffer et al., 1976; Forget et al. 1999; Smith et al., 2001). These processes involve a large fraction of the Martian atmosphere and drive global pressure variations, validating the basic inference that atmospheric activity is linked to the polar-cap behaviour observed in the nineteenth century.